\documentclass[a4paper,fleqn,usenatbib]{mnras}

\usepackage{mathptmx}

\usepackage[T1]{fontenc}
\usepackage{ae,aecompl}

\usepackage{graphicx}	
\usepackage{savesym}
\usepackage{amsmath}	
\savesymbol{iint}
\savesymbol{iiint}
\savesymbol{iiiint}
\savesymbol{idotsint}
\usepackage{txfonts}
\restoresymbol{TXF}{iint}
\restoresymbol{TXF}{iint}
\restoresymbol{TXF}{iiint}
\restoresymbol{TXF}{iiiint}
\restoresymbol{TXF}{idotsint}

\usepackage{pdflscape}
\usepackage{epstopdf}
\usepackage{amssymb}	

\title[Olivine-rich NEAs]{Olivine-rich asteroids in the near-Earth space}

\author[M. Popescu et al.]{
Marcel Popescu$^{1,2}$,\thanks{E-mail:mpopescu@imcce.fr}
D. Perna$^{3,1}$,
M. A. Barucci$^{1}$,
S. Fornasier$^{1}$,
A. Doressoundiram$^{1}$,\newauthor
C. Lantz$^{4,1}$,
F. Merlin$^{1}$,
I. N. Belskaya$^{5,1}$,
and M. Fulchignoni$^{1}$
\\
$^{1}$ LESIA, Observatoire de Paris, PSL Research University, CNRS, Univ. Paris Diderot, Sorbonne Paris \\
Cit\'e, UPMC Univ., Paris 06, Sorbonne Universit\'es,  5 Place J. Janssen, Meudon Pricipal Cedex 92195, France\\
$^{2}$ Astronomical Institute of the Romanian Academy, 5 Cu\c{t}itul de Argint, 040557 Bucharest, Romania\\
$^{3}$ INAF - Osservatorio Astronomico di Roma, Via Frascati 33, 00078 Monte Porzio Catone, Rome, Italy\\
$^{4}$ IAS, UMR 8617, CNRS, Universit\'e Paris-Sud, Bat 121, F-91405 Orsay, France \\
$^{5}$ Institute of Astronomy, V.N. Karazin Kharkiv National University, 35 Sumska Str., 61022 Kharkiv, Ukraine
}

\date{Accepted XXX. Received YYY; in original form ZZZ}

\pubyear{2017}

\begin{document}
\label{firstpage}
\pagerange{\pageref{firstpage}--\pageref{lastpage}}
\maketitle

\begin{abstract}
In the framework of a 30-night spectroscopic survey of small near-Earth asteroids (NEAs) we present new results regarding the identification of olivine-rich objects. The following NEAs were classified as A-type using visible spectra obtained with 3.6 m NTT telescope: (293726) 2007 RQ17, (444584) 2006 UK, 2012 NP, 2014 YS34, 2015 HB117, 2015 LH, 2015 TB179, 2015 TW144. We determined a relative abundance of $5.4\% $ (8 out of 147 observed targets) A-types at hundred meter size range of NEAs population. The ratio is at least five times larger compared with the previously known A-types, which represent less than $\sim1\%$ of NEAs taxonomically classified. By taking into account that part of our targets may not be confirmed as olivine-rich asteroids by their near-infrared spectra, or they can have a nebular origin, our result provides an upper-limit estimation of mantle fragments at size ranges bellow 300m. Our findings are compared with the "battered-to-bits" scenario, claiming that at small sizes the olivine-rich objects should be more abundant when compared with basaltic and iron ones. 
\end{abstract}




\section{Introduction}

The asteroids with olivine-rich compositions provide fundamental clues for understanding the accretion and geochemical evolution of primitive bodies \citep[e.g.][]{2004A&A...422L..59D}. They are expected to be formed \citep[e.g.][]{2014Icar..228..288S} either through magmatic differentiation, being the major constituent of the mantles of most differentiated bodies \citep[e.g.][]{1996M&PS...31..607B}, or through nebular processes which can produce olivine-dominated objects like the R-chondrite parent body \citep{1994Metic..29..275S,2007M&PS...42..155S}.

In the optical region, the reflectance spectra of olivine-rich asteroids show very steep to extremely steep slope at wavelengths short-ward than 0.75 $\mu m$ and a moderately deep absorption feature long-ward \citep{1989aste.conf..298T, 2002Icar..158..106B}. They are classified as A-type \citep[e.g.][]{1983Icar...55..177V,2002Icar..158..146B,2009Icar..202..160D} which is an end-member taxonomic class. 

It was found that these objects have moderately high albedos and a spectral band centered around 1.05 $\mu m$ \citep{2007M&PS...42..155S} which is composed of several absorptions generated by the presence of $Fe^{2+}$ in the olivine structure \citep{1993macf.book.....B}. The 2 $\mu m$ feature characteristic for pyroxene compositions is absent or very weak which suggests almost a pure olivine composition on the surface \citep{2014Icar..228..288S}. Olivine exists also in the S-types asteroid population (associated with olivine-pyroxene assemblages). But, it is a difference between the S-type olivine asteroids and olivine dominated A-type asteroids which reflect real geologic differences in the parent body history and the formation processes \citep{2015Icar..250..623G}.

A very few olivine-rich asteroids (about 30 according to various datasets) which could be representative of the mantle of disrupted differentiated bodies, are known to exist \citep[e.g.][]{DeMeo2018}. This observational result is unexpected when compared with the meteoritic evidences which show that at least 100 chondritic parent bodies in the main belt experienced partial or complete melting and differentiation before being disrupted \citep[][and references there in]{2015aste.book..533S, 2015aste.book..573S} - thus they should have produced a much larger number of olivine - dominated objects. The paucity of these objects in the main belt is called "Missing Mantle Problem" \citep{2015aste.book...13D}. Several scenarios have been proposed as a solution: i) the "battered to bits" scenario \citep{1996M&PS...31..607B} propose that the mantle and crustal material of the original differentiated bodies had been ground down to pieces below the limit of detectability of current spectroscopic surveys; ii) \cite{2011E&PSL.305....1E} proposed that the classic view of asteroids differentiating into a pyroxene-rich crust, olivine-rich mantle, and iron core may be  uncommon. They suggest that only the parent body core heats and melts due to the $~^{26}Al$ decay, but the exterior remains unheated and primitive and it is hiding the evidence of interior differentiation. In addition to these models, there is the problem of multiple scattering of the objects relative to the place where they formed during the early history of Solar System formation \citep{2015aste.book...13D}.

A total of 17 asteroids were identified as A-types during the SMASSII survey \citep{2002Icar..158..146B}. In order to define the limits of this class, they used the five asteroids unambiguously identified by Tholen taxonomy as A-types, namely (246) Asporina, (289) Nenetta, (446) Aeternitas, (863) Benkoela, (2501) Lohja. The SMASSII spectrum of (289) Nenetta shows a subtle absorption feature centered near 0.63 $\mu m$ which may be attributed to various phenomena without obvious relation to the composition \citep{1998JGR...10313675S}.

About 12 near-infrared spectra of A-type asteroids are analyzed by \citet{2014Icar..228..288S}. This sample includes some of the previous objects classified based on visible spectral data. They distinguished two classes, one called monomineralic-olivine which exhibit only 1 $\mu m$ feature and the second for which the spectra exhibit the 1 $\mu m$ feature and a weak 2 $\mu m$ feature corresponding to pyroxene. They did not found any link of the type core-mantel-crust between these objects and asteroids families.

To confirm the A-type candidates from Sloan Digital Sky Survey (SDSS) Moving Object Catalog \cite{2014AAS...22432109D,DeMeo2018} conducted a near-infrared spectral survey of asteroids. They report another 20 A-type asteroids throughout the  Main Asteroid Belt, tripling the number of known A-types. They found that A-types represent only $0.16\%$ of the Main Belt population larger than $\sim1$ km in diameter and they estimated that there are about 600 of these objects  with $H\leq17$, half of which in the absolute magnitude range 16-17. New large spectro-photometric surveys in the near-infrared region \citep{2016A&A...591A.115P} allow to identify new A-type candidates at these size ranges.

Using taxonomic classification of more than 1000 near-Earth asteroids (NEAs) from various sources, \cite{2015aste.book..243B} found that the A-types are uncommon in this population, representing a fraction less than $1\%$. They also noted that  A- and E- types are more common among Mars-crossers ($\approx 5\%$) which can indicate a slow diffusion into Mars-crossing orbits from the Flora or Hungaria regions.

We present new spectroscopic results regarding the identification of olivine rich asteroids in the small NEA's population (absolute magnitudes higher than 20). Due to the fact that these object are faint (typically their visual apparent magnitudes are fainter than 18), the accessible way to recognize them is by means of visible spectroscopy. The spectra discussed in this paper were obtained during a 30-night optical spectroscopic survey of near-Earth asteroids, as part of the NEOShield-2 project \citep{Perna2017}. The results of the observing program allows to estimate the ratio of A-types NEAs smaller than 300 m.

The article is organized as follow: Section 2 describes the observing program and the data reduction performed in order to obtain the spectra. Section 3 introduce the methods used to analyze the asteroid spectra. The probability of associating these asteroids with olivine-dominated compositions is discussed. Section 4 shows the results with respect to the physical and dynamical parameters of each object. The implications of our findings are shown in Section 5. The Conclusions section ends the article. 

\section{Observations and data reduction}

The observations reported in this article were made in the framework of NEOShield-2 project\footnote{\url{http://www.neoshield.eu/}}. The main objectives of the corresponding work-package were to undertake an extensive observational campaign involving complementary techniques with the aim to provide physical and compositional characterization of a large number of NEOs in the hundred-meter size range. We performed a spectroscopic survey at ESO 3.6 m New Technology Telescope (NTT) telescope to obtain information about the composition of the small population of asteroids \citep{Perna2017}. The survey was divided in 12 observing sessions spread almost uniformly (one session every 2-3 months) over two years (04/2015 - 02/2017). Each session consisted in two or three observing nights. This format allowed to cover the majority of asteroids with estimated diameter lower than 300 m and brighter than $V\approx20.5$ (were V is the apparent magnitude) at the observing date. 

The EFOSC2 instrument was used with the Grism $\#1$ diffraction element. A slit width of 2 arcsec was selected considering an average seeing of 1 to 1.5 arcsec. This configuration covers the spectral interval 0.4-0.92 $\mu m$ with a resolution of R$\approx$500. All the observations were made by orienting the slit along the parallactic angle, to minimize the effects of atmospheric differential refraction. The strategy was to observe all asteroids as close to the zenith as possible. The G2V solar analogs were observed in the apparent vicinity and at similar airmasses with the objects of interest. The number of solar analogs observed each night was larger than the number of asteroids with the purpose to avoid any artefacts caused by G2V selection and observation.

The data reduction followed the standard procedures (bias and background subtraction, flat field correction, one-dimensional spectra extraction). We used Octave software to generate IRAF scripts for performing these tasks automatically. The \emph{apall} package was used to extract the one dimensional spectra across the apertures. Wavelength calibration was obtained using emission lines from the He-Ar lamp.  In the final step the reflectivity of the objects was determined by dividing the asteroid spectra with those of the solar analogs. For each asteroid we selected the solar analog which give a result close to the median when comparing with the results obtained with respect to all other solar analogs observed during that night. All spectra were normalized at 0.55 $\mu m$ and the results are presented in Fig.~\ref{fig:Spectra}.

Spectral observations are subject to multiple selection biases due to asteroid properties (size, albedo and distance to observer), and due to equipment \citep{2002aste.book...71J}. Due to the incompleteness of observed versus as-yet undiscovered asteroids at this size range (our survey doubled the number of hundred meter size range asteroids which can be taxonomically classified) any de-biasing technique required to derive statistics between groups with very different properties (e.g. albedos) is poorly constrained. However, because we did a random selection of objects, those with similar sizes and albedos had equal probabilities to be observed.

\begin{figure}
\includegraphics[width=\columnwidth]{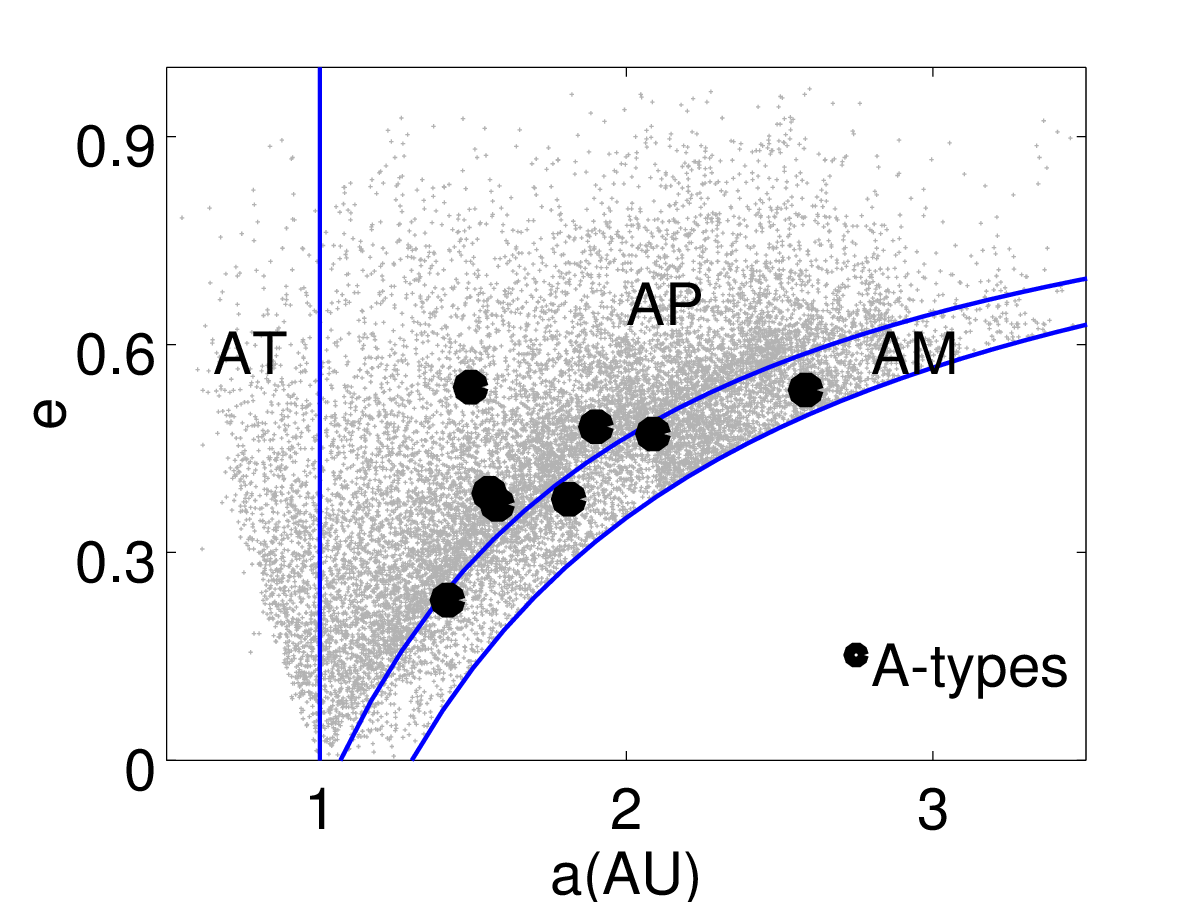}
\caption{The observed A-types asteroids in a ($a,e$) representation of NEAs ($a$ is the semi-major axis and $e$ the eccentricity). The blue lines are the limits of different orbital classes (AM-Amor, AP-Apollo, AT-Aten).}
 \label{fig:ae}
\end{figure}

In this article we focus on the asteroids classified as A-type. This was the case of eight objects, namely: (293726) 2007 RQ17, (444584) 2006 UK , 2012 NP, 2014 YS34, 2015 HB117, 2015 LH, 2015 TB179, and 2015 TW144. Their orbital elements and the observation details are provided in Table~\ref{tab:Circumstances}. The Fig.~\ref{fig:ae} shows that this sample have no particular orbital elements with respect to the known population of NEAs. In addition, the orbital inclination, $i$ of these asteroids is in the $2-8.5^{\circ}$ range. The asteroids (444584) 2006 UK  and 2014 YS34 are catalogued by Minor Planet Center (MPC) as potentially hazardous asteroid (PHA).

\begin{table*}
\centering
\caption{Observing conditions: asteroid designation, the orbit type (AM-Amor, AP- Apollo, *-Potentially Hazardous Asteroid), semi-major axis, eccentricity, inclination, minimum orbital intersection distance with Earth (MOID), mid UTC of the observations, the apparent magnitude (V), the phase angle ($\alpha$), the airmass, the total exposure time (Exp.), and the corresponding solar analog (S.A.) used to get the reflectance, are presented.}
\label{tab:Circumstances}
\begin{tabular}{l c c c c c l c c c l l}
\hline
Designation &Orbit&a(AU)   &e    &i($^\circ$) &MOID(AU) &UTC              &Vmag &$\alpha$ ($^\circ$) & Airmass &Exp.(sec) &S.A.\\
\hline
(293726) 2007 RQ17&AP   &1.579 &0.37 &2.0         &0.021  &2015-06-08T07:44 &18.1 &51                &1.288    &2x600   &SA 107998\\
(444584) 2006 UK &AP*  &1.493 &0.54 &4.7         &0.014  &2016-05-12T06:28 &16.5 &29                &1.731    &1x600   &SA 107998 \\
2012 NP     &AM   &2.088 &0.47 &7.4         &0.098  &2015-07-20T02:17 &18.9 &41                &1.356    &2x900   &HD 111244 \\
2014 YS34   &AP*  &1.553 &0.39 &6.9         &0.007  &2015-06-09T07:14 &20.0 &20                &1.249    &2x1125  &SA 1021081 \\
2015 HB117  &AM   &1.417 &0.23 &6.5         &0.081  &2015-06-08T04:24 &20.2 &33                &1.359    &2x1200  &SA 1021081 \\
2015 LH     &AP   &1.902 &0.48 &3.9         &0.003  &2015-06-09T06:30 &19.1 &44                &1.426    &1x180   &SA 1021081 \\
2015 TB179  &AM   &2.586 &0.53 &8.1         &0.226  &2015-11-05T07:34 &20.0 &27                &1.203    &3x1200  &HD 11123 \\
2015 TW144  &AM   &1.813 &0.38 &8.4         &0.201  &2015-11-05T05:46 &19.3 &22                &1.285    &2x900   &Hyades64 \\
\hline
\end{tabular}    
\end{table*}

\section{Methods used to analyze data}\label{MET}

The A-type is a distinct end-member class in most of the recent taxonomies \citep[e.g.][]{1984atca.book.....T, 1987Icar...72..304B, 1989aste.conf..298T, 2000Icar..146..204F, 2002Icar..158..146B, 2009Icar..202..160D}. Based on the spectra obtained over 0.44 - 0.92 $\mu m$ spectral interval, \citet{2002Icar..158..146B} provides the following definition "Very steep to extremely steep UV slope short-ward of 0.75 $\mu m$, and a moderately deep absorption feature, long-ward of 0.75 $\mu m$. A subtle absorption feature is often present around 0.63 $\mu m$".

Several authors \citep[e.g.][]{2015Icar..250..623G} argue that some of the A-types identified based on visible spectra  does not show near-infrared data compatible with olivine dominated asteroids.  We note that from 17 objects reported in SMASSII survey \citep{2002Icar..158..106B} as belonging to A-type asteroids, 13 have near-infrared data up to 2.45 $\mu m$. In this sample, six objects are classified as A type: (246) Asporina, (289) Nenetta, (446) Aeternitas, (863) Benkoela, (1951) Lick and (2501) Lohja \citep{2009Icar..202..160D,2004A&A...422L..59D}. The asteroid (4142) Dersu-Uzala is reported as A-type by \citet{2004P&SS...52..291B} but displays a 2 $\mu m$ feature which indicates a non-negligible pyroxene content. Different classifications are provided for (3352) McAuliffe - \citet{2004Icar..170..259B} reported it as A-type, and \citet{2001PhDT.......121W} as S-type.  The asteroids (1126) Otero, (2715) Mielikki, (4713) Steel, and (5641) McCleese are classified as Sw-type and (2732) Witt as L-type by \citet{2009Icar..202..160D}. Although this statistics is extremely poor, it provides a rough estimation of the accuracy for recognizing olivine dominated assemblages based on visible data.

Within the limits allowed by our data (spectral interval and signal to noise ratio), we used several methods to ensure the correct classification as A-type for the asteroids shown in this paper: i) we computed the mean square differences with respect to all taxonomic types defined by \citet{2002Icar..158..146B}; ii) we determined the position of the maximum, the slopes in the (0.45 - 0.65) $\mu m$ (denoted as $BR_{slope}$) and (0.75-0.92) $\mu m$ (denoted as $IZ_{slope}$) spectral intervals (Table~\ref{tab:Prop}) and we compared them with those of standard taxonomic types; iii) we compared these spectral parameters with those of known A-types; and iv) we compared our data with all spectra available in the RELAB database (NASA Reflectance Experiment Laboratory\footnote{\url{http://www.planetary.brown.edu/relabdocs/relab.htm}}). These methods are presented bellow.

\emph{i)} The first two spectral classes from Bus taxonomy that matched our spectra according to the mean square differences are presented in Table~\ref{tab:TaxMSq} and plotted in Fig.~\ref{fig:SpectraTaxa}. All objects are classified as A-type considering a calculated distance at half compared with the second match which is Sa type for the majority. This is fully in agreement with an olivine dominated spectrum.

\begin{table}
\centering
\caption{The two spectral classes from Bus taxonomy that best matched our spectra and their mean square differences (msq).}
\label{tab:TaxMSq}
\begin{tabular}{l c c c c}
\hline
Designation &Tax 1 &msq.tax1&Tax 2 &msq.tax2\\
\hline
(293726) 2007 RQ17   &A&0.0109&Ld&0.0153\\
(444584) 2006 UK   &A&0.0007&Sa&0.0012\\
2012 NP     &A&0.0024&Sa&0.0049\\
2014 YS34   &A&0.0008&Sa&0.0024\\
2015 HB117  &A&0.0015&Ld&0.0024\\
2015 LH     &A&0.0060&Sa&0.0102\\
2015 TB179  &A&0.0007&Sa&0.0015\\
2015 TW144  &A&0.0016&Sa&0.0030\\
\hline
\end{tabular}    
\end{table}

\begin{table*}
\centering
\caption{Some characteristics of the NEAs studied in this article. The asteroid designations, absolute magnitude, estimated diameter, and the spectral slopes ($BR_{slope} (\%/0.1~\mu m)$, $IZ_{slope}(\%/0.1~\mu m)$) computed based on our observation.}
\label{tab:Prop}
\begin{tabular}{l c c c c c c c}
\hline
Designation &H(mag.) &Diam.(m) &Max.pos($\mu m$)& $BR_{slope}(\%/0.1~\mu m)$ &$Err_{BRslope}$ & $IZ_{slope}(\%/0.1~\mu m)$ &$Err_{IZslope}$ \\
\hline
(293726) 2007 RQ17 &22.5    &96       &0.762           &40.2          &0.5             &-13.5         &1.3\\
(444584) 2006 UK &20.2    &277      &0.775           &25.8          &1.0             &-13.4         &3.5\\
2012 NP     &21.3    &167      &0.763           &29.8          &0.5             &-14.9         &2.0\\
2014 YS34   &20.8    &210      &0.775           &25.0          &1.0             &-10.0         &5.1\\
2015 HB117  &23.6    &58       &0.762           &25.8          &1.0             &-6.0          &3.0\\
2015 LH     &27.3    &11       &0.736           &37.9          &1.7             &-21.1         &3.8\\
2015 TB179  &20.6    &231      &0.748           &24.7          &0.8             &-12.7         &2.3\\
2015 TW144  &20.7    &220      &0.738           &26.3          &0.6             &-12.3         &1.3\\
\hline
\end{tabular}    
\end{table*}

\emph{ii)} The position of the peak of our identified spectra (Table~\ref{tab:Prop}) ranges from 0.736 $\mu m$ (which belongs to 2015 LH) to 0.775 $\mu m$ (which belongs to 2014 YS34) with an average value $\lambda_{peak} = 0.754\pm0.017$ $\mu m$ . The $BR_{slope}$ computed over the interval (0.45 - 0.65) $\mu m$ is for the majority larger than 25$\%/0.1~\mu m$. These values compare only with the A type objects that were classified by SMASS and which have the average $BR_{slope}^{SMASS} = 25.1\pm3.5\%/0.1~\mu m$. For comparison, the S-complex objects reported by \citet{2009Icar..202..160D} have $BR_{slope}^{Scomp} = 14.3\pm3.8\%/0.1~\mu m$.

The presence of the band at 1 $\mu m$ is critical to differentiate between the A-types and other possible red type spectra corresponding to L, Ld, and D types from Bus taxonomy. The computed values for the $IZ_{slopes}<-6\%/0.1~\mu m$ strongly indicates the presence of the feature even for low signal to noise ratio spectra. All this parameters point to an A-type classification, well distinctive from L/Ld type or from the most common S-type ( Fig.~\ref{fig:SpectraTaxa}).

\emph{iii)} Depending on the shape of the maximum, \citet{2002Icar..158..146B} identified two spectral curves associated with the A-class.
The first one, which is similar with the spectrum of (289) Nenetta and (863) Benkoela, has the maximum at about $\approx$ 0.75 $\mu m$  sharply peaked with 1 $\mu m$ band particularly rounded while the second has a broader reflectance like the spectrum of (246) Asporina and no upward curvature out to 0.92 $\mu m$. To asses a comparison we computed the mean square error differences in the (0.70, 0.80) $\mu m$ spectral interval between our asteroids spectra and the SMASSII spectra \citep{2002Icar..158..106B} of (289) Nenetta, (863) Benkoela and  (246) Asporina. Within the limit of noise, we found that (293726) 2007 RQ17 and 2015 LH have more (863) Benkoela like spectrum and (444584) 2006 UK, 2012 NP, 2014 YS34, 2015 HB117, 2015 TB179, and 2015 TW144 have a bowl shape maximum similar with (246) Asporina. However, this is a rough comparison as most of the spectra show intermediate shapes between these references.

\emph{iv)} The comparison with RELAB spectral database considering only the visible spectra provides few constraints with respect to the composition of the object as multiple different solutions may be found to match the same visible data. Also, it has to be noted that space-weathering effects change the spectral curve of asteroids \citep{2006Icar..184..327B}. However, the comparison deserves to be performed as a measure of the various solutions which can be found. The most relevant results of this comparison are shown in Table~\ref{tab:RelabComparison} and plotted in Fig~\ref{fig:Relab}. The comparison was performed using M4AST\footnote{\url{http://m4ast.imcce.fr/}} tool \citep{2012A&A...544A.130P}.  The solutions show that these spectral data match the different mixtures of olivine or thin sections of olivine from the meteorites. In particular one of the RELAB spectrum that fit our observed data was of a particulate olivine fayalite mineral. Nevertheless, among spectral curve matches there are also several corresponding to typically ordinary chondrite meteorites (e.g. L4 ordinary chondrite meteorite Barratta).
                                                                                                                                                                                                                                                                                                                                                                                                                                                                                                                                                                                                                                                                                                                                                                                                                           
\begin{equation}
D = \frac{1329}{\sqrt{p_V}}10^{-0.2H}
\label{diam}
\end{equation}

The size of the asteroids presented in this paper was estimated by computing their equivalent diameter using the absolute magnitude H reported on the IAU Minor Planet Center website\footnote{\url{http://www.minorplanetcenter.net}} and the average albedo for A-types $p_v = 0.19$ \citep{2011ApJ...741...90M} using the Eq.~\ref{diam} \cite{2007Icar..190..250P}). The values are reported in Table~\ref{tab:Prop}.


\section{Results}
\begin{figure*}
\begin{center}
\includegraphics[width=5.7cm]{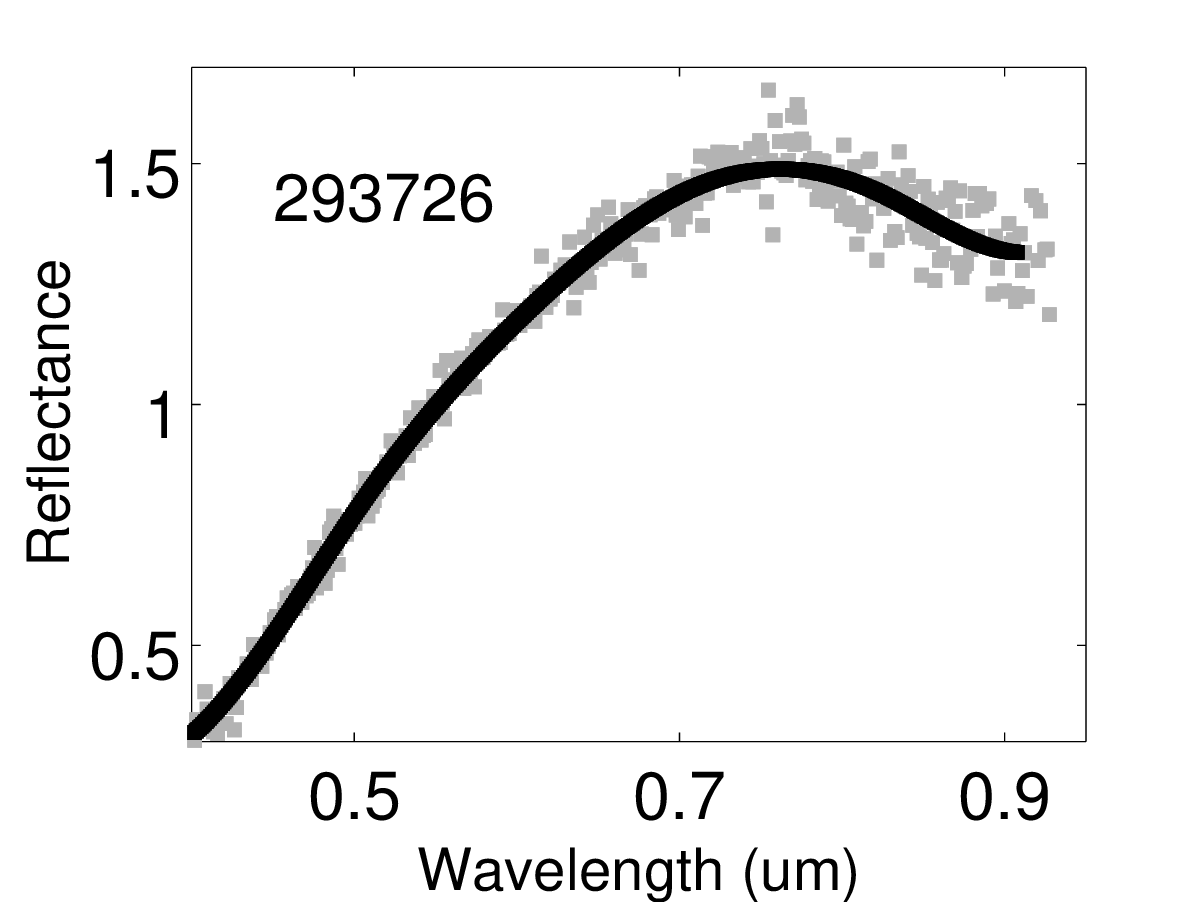}
\includegraphics[width=5.7cm]{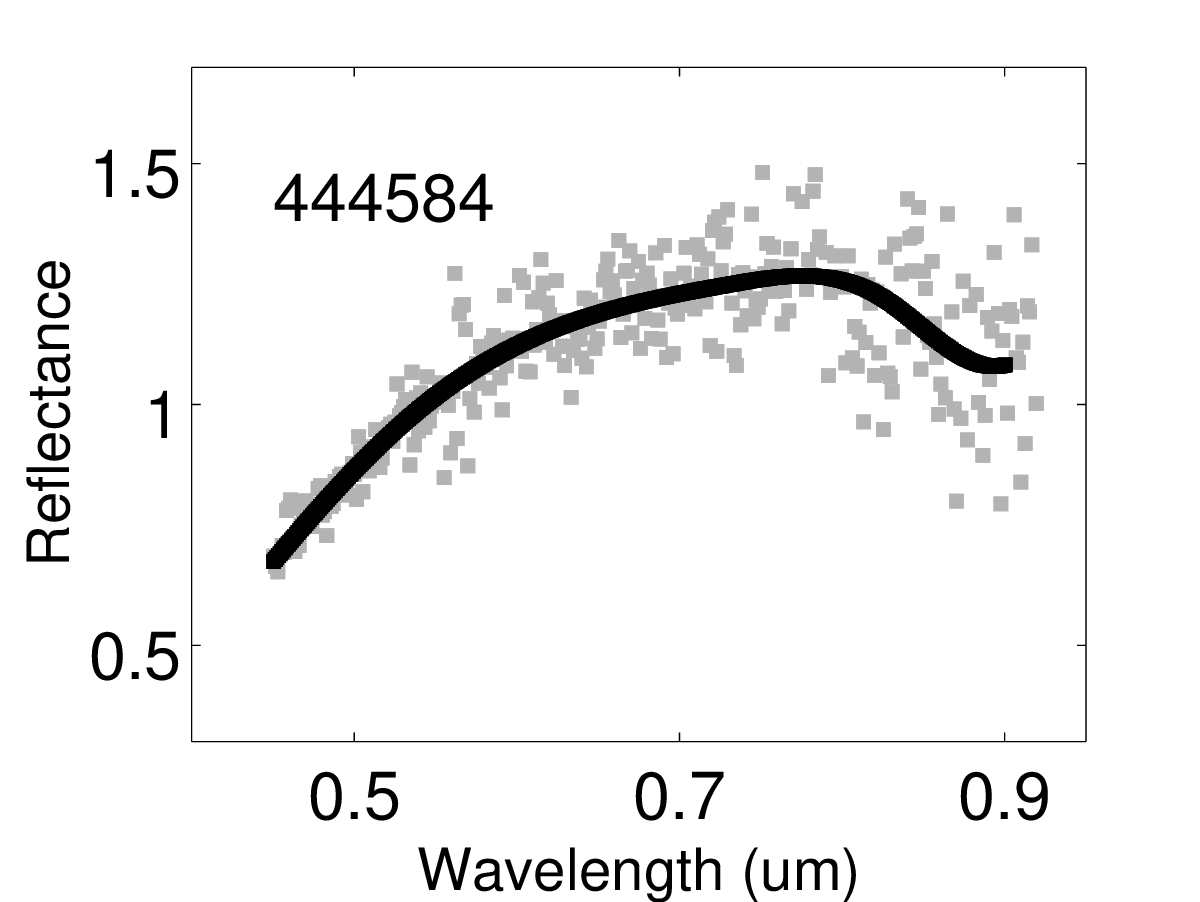}
\includegraphics[width=5.7cm]{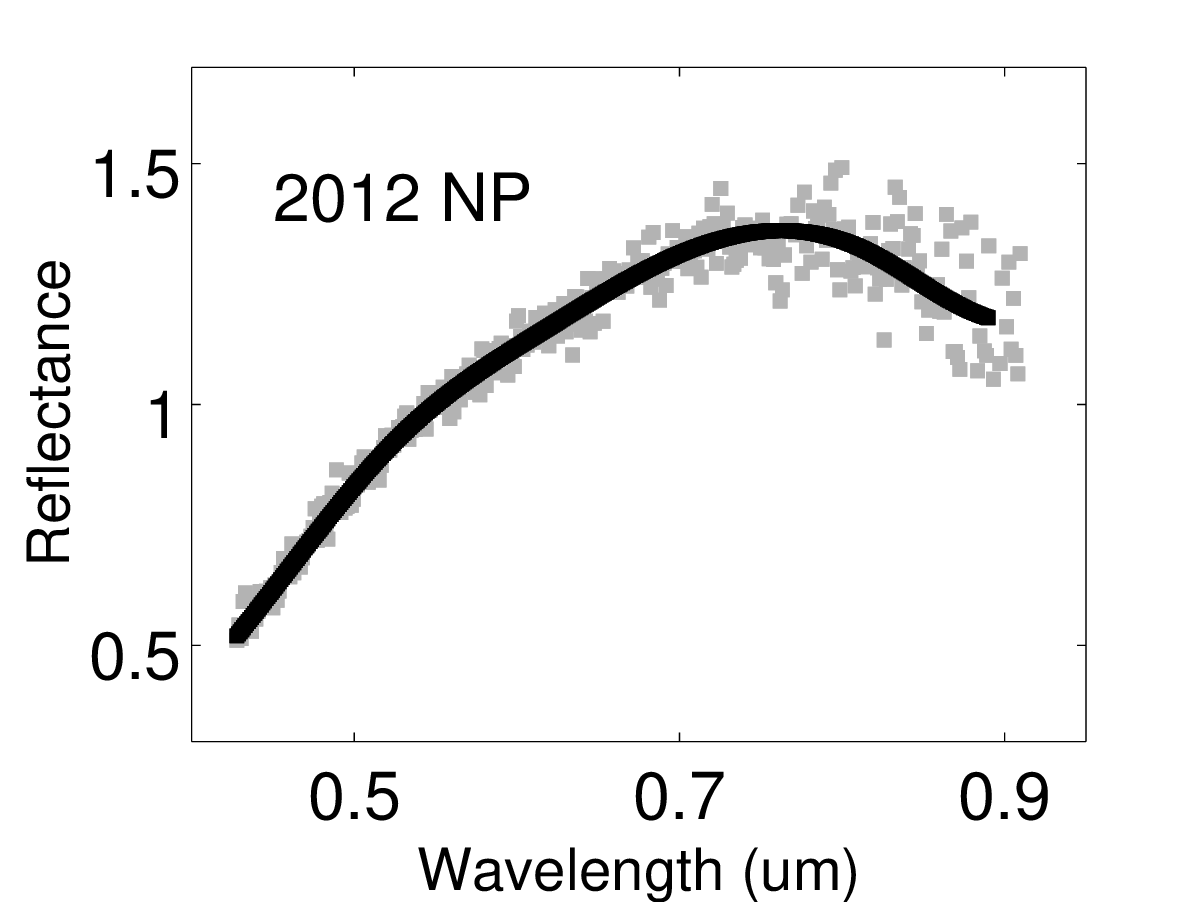}
\includegraphics[width=5.7cm]{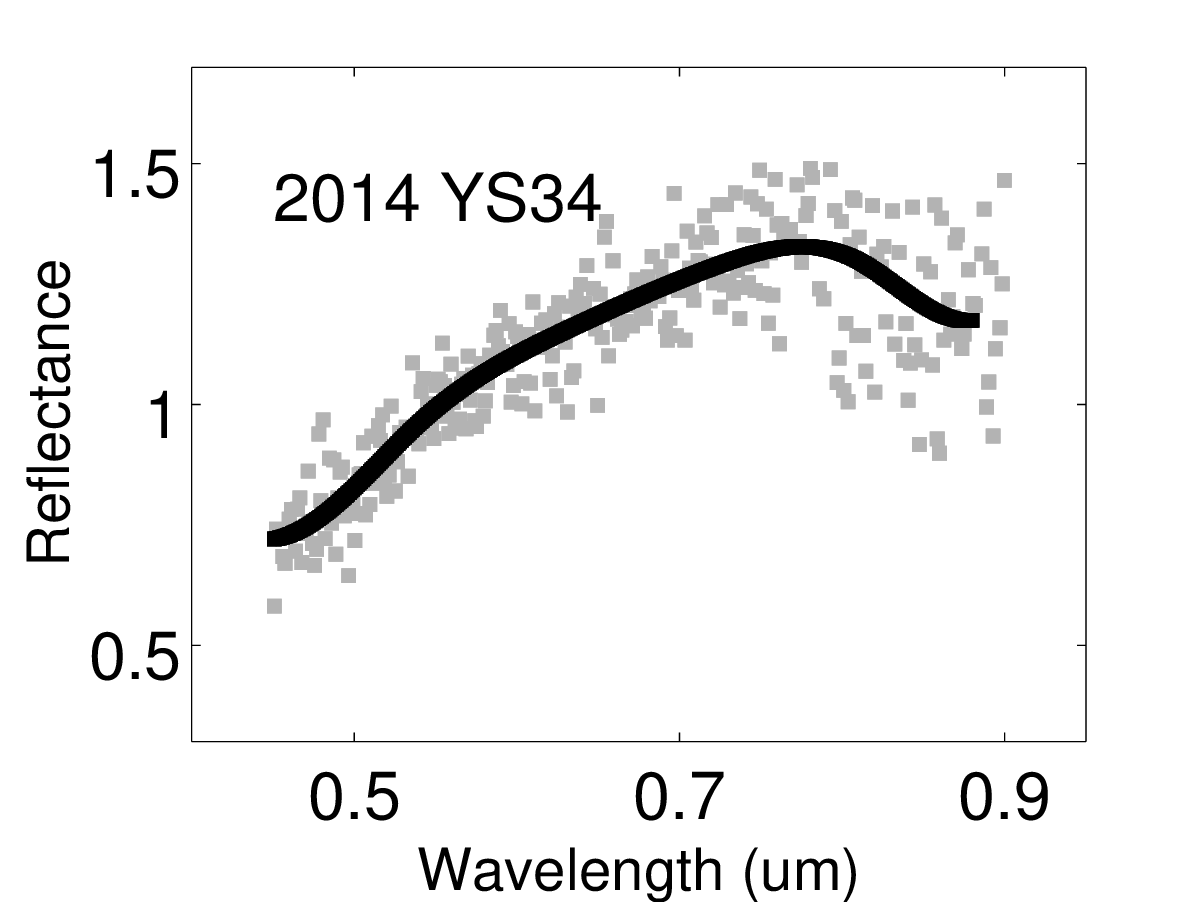}
\includegraphics[width=5.7cm]{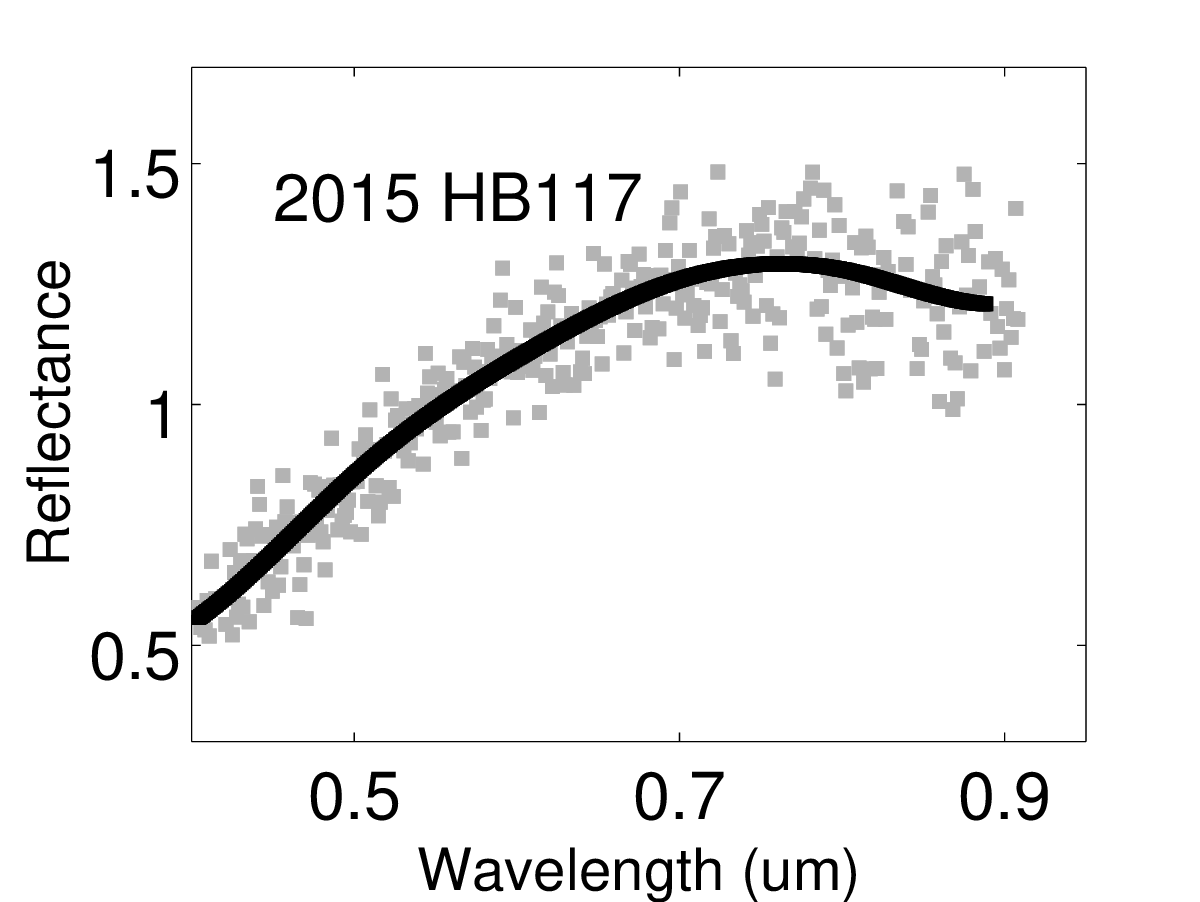}
\includegraphics[width=5.7cm]{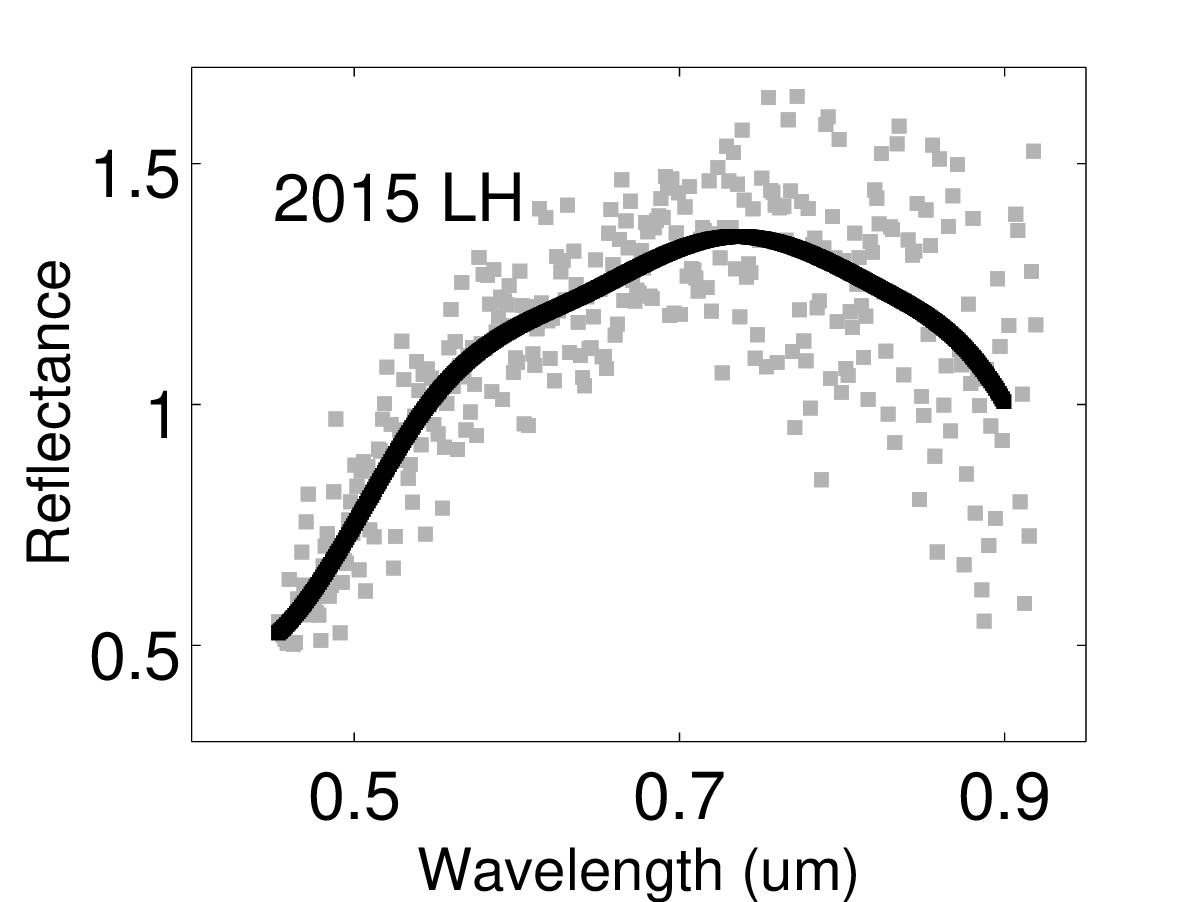}
\includegraphics[width=5.7cm]{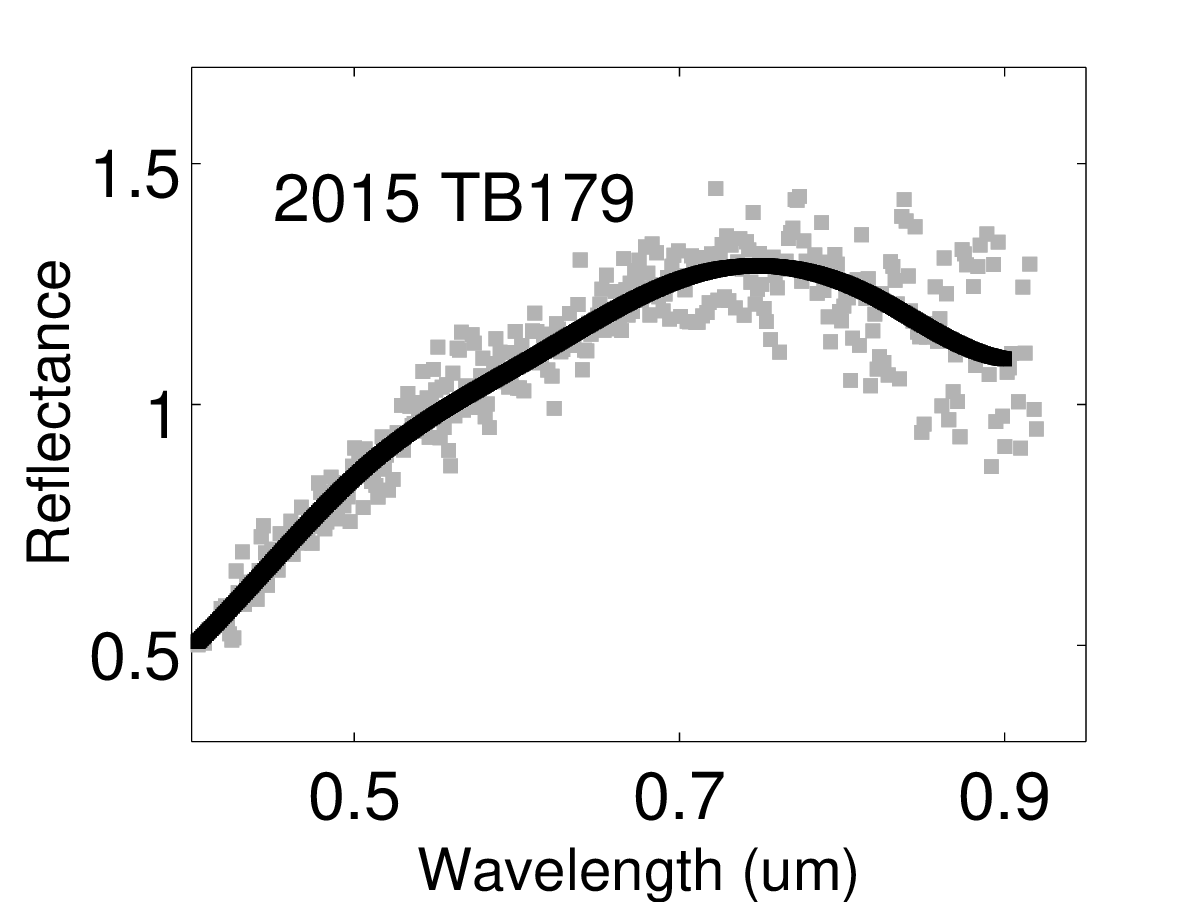}
\includegraphics[width=5.7cm]{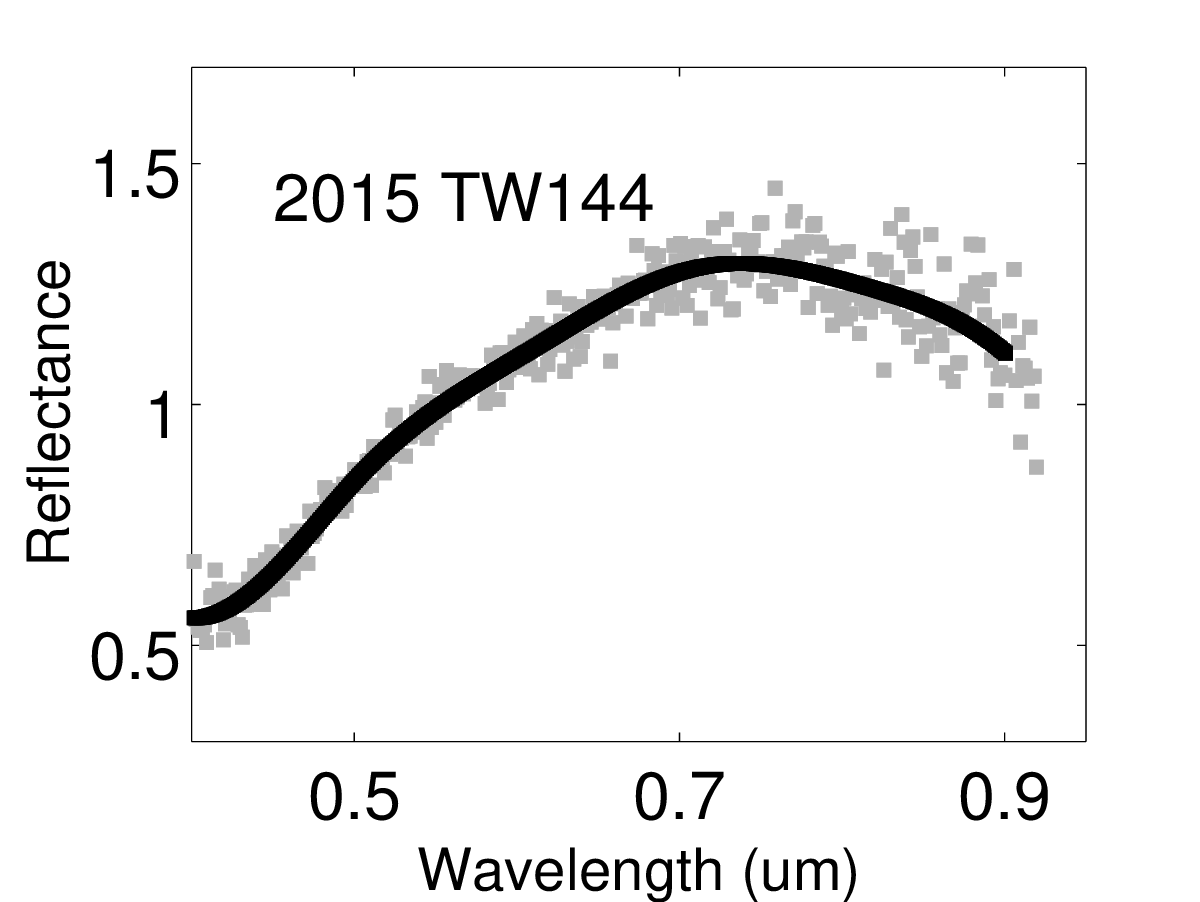}

\end{center}
\caption{Visible spectra of (293726) 2007 RQ17, (444584) 2006 UK, 2012 NP, 2014 YS34, 2015 HB117, 2015 LH, 2015 TB179, 2015 TW144. All spectra are normalized to 0.55 $\mu m$. A spline fit of the data is shown with black while the values extracted from the images are shown in grey.}
\label{fig:Spectra}
\end{figure*}

From the total sample of 147 observed near-Earth asteroids with an absolute magnitude H$\leq20 $ mag we found  eight asteroids classified as A-type. Their spectra are plotted in Fig.~\ref{fig:Spectra}. Three of these objects have an estimated diameter less than 100 m - (293726) 2007 RQ17, 2015 HB117, and 2015 LH while the other five are about $\sim 200$ m size. Table~\ref{tab:Prop} summarizes the diameters and the computed spectral parameters. In the public databases there is no other physical information available, except the orbital parameters and the absolute magnitude derived from astrometric data (form Minor Planet Center website). Part of these objects have sufficient astrometric data for follow up at future Earth close approaches. But the smallest objects, such as 2015 LH, were observed over a very short orbital arc and their orbit uncertainty is high. Thus, they require astrometric recovering \citep[e.g.][]{2018A&A...609A.105V}.

\emph{(293726) 2007 RQ17} is an Apollo type asteroid for which we estimated a diameter of 96 m. It was discovered in September 2007, and although it is a hundred meter size object, it has sufficient astrometric observations to receive a permanent designation (i.e. a number). It has a perihelion  at $q = 0.996$ AU and the MOID (minimum orbital intersection distance) with Earth 0.021 AU.  This asteroid has the reddest spectrum among our sample - characterized by the largest $BR_{slope} $ of about $40.2 \%/0.1~\mu m$. Our observations were performed at a phase angle of $\alpha = 51^{\circ}$, thus a phase reddening factor may account for this high slope \citep{Perna2017}. No spectral feature can be distinguished below 0.7 $\mu m$.  The mineral spectrum matching the observed data is the one of an olivine fayalite sample.

\emph{(444584) 2006 UK} is PHA  with an estimated diameter of 277 m and it is the largest object from the sample reported here. It has a MOID with Earth of 0.014 AU and it is on a eccentric orbit ($e = 0.54$), with a perihelion $q = 0.689 AU$. It has a broad maximum at $0.775~\mu m$ and its spectral curve is the closest to the A-type template. The spectrum shows a change of slope around 0.65 $\mu m$, which may suggest a possible feature at these wavelengths.

\emph{2012 NP} is an Amor type object with an equivalent diameter of 167 m. It was observed at $\alpha = 41^{\circ}$ and it has a $BR_{slope} $ of about $29.8 \%/0.1~\mu m$, thus being the second reddest object. This is in agreement with the correlation between $BR_{slope}$ and phase angle for the A-types, outlined by \cite{Perna2017}.

\emph{2014 YS34} is an Apollo type asteroid with an equivalent diameter of 210 m. It is catalogued as PHA. The spectrum shows a slightly change of slope around 0.65 $\mu m$ which can indicate a possible feature. For this object MPC records astrometric positions covering an arc-length over 888 days, thus allowing follow up observations.

\emph{2015 HB117} is an Amor type object with an estimated diameter of 58 m. Astrometric positions are reported over 83 days. The feature at 0.9 $\mu m$ is less pronounced compared with the other objects which is quantified by the small value $ IZ_{slope} = -6\pm3\%/0.1~\mu m$. Due to this fact, a classification as an L-type can not be totally excluded (being at the limit of noise).

\emph{2015 LH} is the smallest object from the observed sample, with an estimated diameter of 11 m. Its observational arc covers only 5 days. Our spectroscopic observations were performed when this asteroid was at $0.011$ AU geocentric distance. Although this spectrum has the lowest signal to noise ratio ($\approx 11$), its $BR_{slope} $ is fully compatible with A-type classification.  It shows a change of slope around $0.65~\mu m$. The shape of the spline-fit indicate a feature at these wavelengths.

\emph{2015 TB179} is a 231 m diameter Amor type object. It has the largest semi-major axis $a = 2.586$ AU and its aphelion $q = 3.968 AU$ in the outer part of the Main Belt. The spectrum is well-consistent with the mean A-type spectrum.

\emph{2015 TW144} is a 220 m Amor type object. Its spectrum shows a broad peak at 0.738 $\mu m$ and a weak feature around 0.45 $\mu m$.

\section{Discussions}

Within the NEAs population only two asteroids have been recognized as olivine dominated NEAs \citep{2014Icar..228..288S}, namely (1951) Lick \citep{2004A&A...422L..59D} - a 5.57 km object which belongs to Mars-crossing asteroid population, but its orbital elements (i.e. perihelion distance q = 1.305 AU) are close to borderline of NEA's (i.e. $q\leq1.3$); and (136617) 1994 CC \cite{2011P&SS...59..772R} which is a triplet NEA with a diameter of 650 m for the primary, 10 m for the secondary, and 5 m of the tertiary component (IAUC 9053).  

By using visible and near-infrared data, \cite{2004A&A...422L..59D} found for (1951) Lick a surface composed by almost pure olivine, estimating the Fosterite content to be $Fo = 90\pm10\%$ (low-iron content). Their result is comparable with the one reported by \citep{2014Icar..228..288S} who found $Fo = 70\pm 5\%$. Based on near-infrared data acquired with SpeX at IRTF telescope, \cite{2011P&SS...59..772R} found that (136617) 1994 CC has Mg-rich ($Fo = 90$) olivine composition, similar to Mg-rich pallasites with a low metal component. 

In the existing spectral databases and those of physical properties of asteroids there are very few other NEAs reported to have a taxonomic type compatible with olivine rich composition. We mention only those that are not ambiguously classified (inconsistent classification when considering data reported by different authors). Thus, based only on spectral data in the visible region (471240) 2011 BT15 \citep{2014PASJ...66...51K} and (488515) 2001 FE90 \citep{2009ATel.2116....1H} are reported as A-types. Both objects have an equivalent diameter in the range of 200-300 m (H = 21.7 mag for 2011 BT15 and H = 20.1 mag for 2001 FE90). The next type which may correspond to an olivine rich composition is Sa. The objects  (136849) 1998 CS1, (275677) 2000 RS11 and 1993 TQ2 (see EARN database \footnote{\url{http://earn.dlr.de/nea/}} and corresponding references for a detailed description) are classified as Sa. \citet{2016Icar..268..340C} report an A-type object within a sample of 230 NEAs based on SDSS data. Overall, we can summarize that from the total number of NEAs spectrally characterized up to now \citep[more than 1\,000 according to][and about 735 in the EARN database]{2015aste.book..243B} less  than $1\%$ have spectra compatible with an olivine rich composition.

The ratio of A-types can be outlined relative to S-complex asteroids which have similar albedo values. The S-types, which show spectra similar with ordinary chondrite material, represent more than half of all measured NEAs \citep{2015aste.book..243B}. Thus, it can be concluded to a fraction less than $2\%$ of olivine dominated achondrite asteroids compared to those associated with ordinary chondrite assemblages. Moreover, the percentage of A-types NEAs is one order less when compared with basaltic asteroids which are considered to originate in the crust of differentiated bodies.

We note that with the exception of Lick which is 5.57 km diameter, the other A-types NEAs are in the hundred meter size range. This implies a very small amount of pure olivine material in terms of volume and mass - considering an average density of $3.73\pm1.40\frac{g}{cm^3}$ \citep{2012P&SS...73...98C}.

Compared with these results, we found a fraction of $\approx(5.4\%)$ olivine dominated asteroids within our observed sample of 147 objects with diameter less than 300 m. This result raise several questions: what are the reasons for finding more A-types within the small NEAs population? is there a difference when comparing the small size objects with the larger ones? what is the origin of these small olivine-dominated objects? 

Due to the fact that only visible data is used to classify them as A-type and the statistically low number of observed objects, several debates can be addressed concerning both the observational program and the accuracy of the classification. These are discussed bellow:
\begin{itemize}
 \item Risk of observational bias. By considering the ratio of A-type objects over the commonly S-types, which have similar albedo, it can be avoided the bias introduced due to the tendency to observe brightest objects when considering the same size. The ratio of S-complex asteroids within the NEOShield-2 observed sample is $\approx54\%$ \citep{Perna2017} comparable with the one reported by \cite{2015aste.book..243B}. Thus, our finding imply a relative fraction of A-type over S-types for NEAs lower than 300 m of $10.3\%$ (8 A-types and  78 S-complex objects).
 
 From the dynamical point of view, as shown in Fig.~\ref{fig:ae} and Table~\ref{tab:Circumstances}, their orbital elements are common for the near-Earth objects and there is no grouping in (a,e) orbital elements space. Nevertheless, we note that all objects have low inclinations ($2-8.5^\circ$), which may point to an origin close to the reference plane. 
  \item Misclassification due to space-weathering and/or phase-angle effects. These effects may account for the reddening of spectra, making a common S-type with ordinary chondrite composition to be confused with an olivine dominated object. Available experimental and numerical modeling of the weathering effect on olivine-rich surfaces has shown an increase of visible slope together with considerable reduction of the 1 micron absorption band \citep[e.g.][]{2014Icar..237...75K}. \citet{2012Icar..220...36S} showed that the increase of spectral slope caused by weathering and phase reddening is of the same value. These effects can lead to ambiguous classifications only for objects close to the boundaries between taxonomic classes \citep{2012Icar..220...36S}. In our case the phase reddening trend is well-seen and constitutes about $0.5\%/^\circ$ \citep{Perna2017}. The reddening effect does not influence the classification of our objects observed in the range of phase angles of 20-51$^\circ$. 
 \item Ambiguous classification due to lack of the near-infrared data. The A-types identified based on visible spectra must be considered cautiously as their near-infrared data may infirm this classification \citep[e.g.][]{2015Icar..250..623G}. Indeed, the A-types are confirmed by the lack of 2 $\mu m$ band. But as shown in the Methods section, although there is a very low number of objects know, it can be infer that $\sim$half of the asteroids classified as A-type are found to have olivine-dominated composition.
 \item Systematic errors introduced by the instrument or by data reduction pipeline. A complete description of the sample of NEAs observed NEOShield-2 survey is presented by \citet{Perna2017}. Because the slope is one of the main parameters that differentiates between the A-type and the S-complex objects we verified if it was affected by a systematic error. The fact that the average slope for S-complex objects observed by NEOShield-2 has a value of $BR_{slope}^{NS} = 13.5\pm3\%/0.1~\mu m$ comparable with $BR_{slope}^{Scomp} = 14.3\pm3.8\%/0.1~\mu m$ computed for the sample reported by \citet{2009Icar..202..160D}, gives confidence in our findings.
 
 We searched for other spectral observations of the objects reported in NEOShield-2 sample. We found nine objects with spectra reported by SMASS-NEO program \footnote{\url{http://smass.mit.edu/minus.html}}. Within the limits of noise, the results are consistent and no systematic difference has been identified. 
 
\end{itemize}

Olivine can be formed through different mechanism: accretion of grains from an oxidized nebular region without significant post-accretionary heating (R chondrites), partial melting with extraction of a basaltic melt leaving an olivine-rich residue (brachinites), or crystallization of olivine from a melt to form an olivine mantle or olivine-rich layer \citep{1998JGR...10313675S}. Small olivine objects should be common, at least as a result of collisions of the hypothesized large number of primordial  differentiated objects (see the Introduction section). In the "battered to bits" scenario \citep[][and references in]{1996M&PS...31..607B}, the small size of these objects is considered as an explanation for the lack of observational data of olivine-dominated compositions.

Olivine spectral features are known to be greatly reduced in contrast by mixture with pyroxenes and other Fe-bearing mineral phases \citep{1981JGR....86.7967S}. Hence the identification of olivine-dominated A-type asteroids is expected to be easier for smaller bodies, as they more probably include more concentrated deposits of purest olivine phases.

NEAs are an unstable population, with lifetime in the order of millions of years \citep[e.g.][]{1996EM&P...72..133M,1997Sci...277..197G,2002Icar..156..399B}. Dynamical models predict that most of these objects come from the inner ($\sim61\%$), middle ($\sim24\%$) and outer ($\sim8\%$) parts of the Main Belt \citep{2002Icar..156..399B}. Due to their large eccentricities they have a short collision lifetime \citep[e.g.][]{1993LPI....24..159B}. Thus, one of the plausible scenario for the origin of small A-types reported here is the violent collisions of differentiated bodies resulting in fragments of the mantle of hundred meter size range.

Nevertheless, the remnants of collisions of differentiated asteroids should be in a greater volume compared with the iron and basaltic asteroids. But, even if all our eight possible A-types ($\sim5.4\%$) are confirmed as olivine-dominated composition, their percentage barely compares with those of basaltic asteroids in the NEA's population, and it is much less than expected from the model of a differentiated object where mantle material is more abundant than iron core and basaltic crust.

We note that, although the eccentricity and semi-major axis of the observed A-types are very different, their inclinations are in tight range of $2-8^\circ$. Thus, a common origin of these eight possible olivine-dominated objects can not be totally excluded. Such compositional grouping of A-type asteroids has been recently shown. The largest members of Eureka family of Mars Trojan asteroids are olivine dominated objects that share a common origin \citep{2017MNRAS.466..489B, 2017Icar..293..243C, 2017NatAs...1E.179P}. Using numeric simulation, \cite{2017NatAs...1E.179P} show that Mars Trojans are more likely to be impact ejecta from Mars and that olivine rich asteroids in the Hungaria population may have the same martian origin.  

\section{Conclusions}

We found eight asteroids classified as A-type in a sample of 147 NEAs spectrally observed in the optical region \citep{Perna2017}. The A-type asteroids are associated with olivine rich compositions. This result allows to estimate a fraction of $\sim5.4\%$ of olivine dominated NEAs for objects with sizes less than 300 m. This fraction is more than five times larger compared with ratio of known A-type asteroids in the NEAs population.

Although the existence of these objects involve violent collisions of large differentiated bodies, as considered by "battered to bits" scenario, this fraction is far less than expected to result from the predicted number of primordial differentiated objects. In the hypothesis that NEAs reflect the broad compositional distribution of Main Belt, this result of $\sim5.4\%$  gives an upper limit for the number of olivine dominated asteroids at hundred meter size range.

The result presented in this paper follows the premise that the makeup of NEO population is heavily affected by asteroid disruption events that create many small fragments that then move swiftly toward the resonances into NEO space. There may not be evidence of older breakups because those smaller fragments have cleared out of both the main belt and the NEO population long ago.

\section*{Acknowledgements}

This work is based on observations collected at the European Organisation for Astronomical Research in the Southern Hemisphere under ESO programme 095.C-0087. We acknowledge financial support from the NEOShield-2 project, funded by the European Union's Horizon 2020 research and innovation programme (contract No. PROTEC-2-2014-640351). DP has received further funding from the Horizon 2020 programme also under the Marie Sklodowska-Curie grant agreement n. 664931. MP received partial funding from a grant of the Romanian National Authority for Scientific Research and Innovation, CNCS - UEFISCDI, project number PN-II-RU-TE-2014-4-2199.  This work was supported by the Programme National de Planétologie(PNP) of CNRS/INSU, co-funded by CNES.
The article make use of data published by the following web-sites Minor Planet Center, E.A.R.N- The Near-Earth Asteroids Data Base,  SMASS - Planetary Spectroscopy at MIT, and RELAB Spectral Database.
We want to specially thank to F.~E. DeMeo for the constructive and helpful suggestions.




\bibliographystyle{mnras}
\bibliography{Atype.bib} 



\appendix
\clearpage
\newpage
\onecolumn
\section{Spectral comparison with data available in RELAB database}

\begin{table*}
\centering
\caption{Comparison with RELAB database. The SampleID of the most relevant spectrum is given. Additional information including sample name, type, sub-type and texture are given as they are provided by the RELAB database. The spectra are shown in Fig.~\ref{fig:Relab}}
\label{tab:RelabComparison}
\begin{tabular}{l l l l l l l} 
\hline     
Asteroid   &SampleID     &Sample Name          &General Type &Type               &Sub Type                        &Texture\\
\hline
293726     &DH-MBW-006   &Fayalite $1-30\mu m$ &Mineral      &Silicate(Neso)     &Olivine Fayalite                &Particulate ($\leq30\mu m$)   \\
444584     &MH-CMP-002   &Barratta             &Rock         &Ordinary Chondrite &L4                              &Slab                           \\
2012 NP    &DD-MDD-045   &Fo 10                &Mineral      &Silicate (Neso)    &Olivine                         &Particulate($\leq3<45\mu m$)   \\
2014 YS34  &LM-LAM-026   &Dhajala              &Rock         &Ordinary Chondrite &H3-4 Olivine-Bronzite           &Thin Section                   \\
2015 HB117 &OM-PCP-001   &01OP1c               &Rock         &Ultramafic         &Dunite                          &Particulate ($250-1000\mu m$)  \\
2015 LH    &PO-CMP-070   &St.Peter's Fayalite  &Mineral      &Silicate (Neso)    &Olivine Fayalite                &Particulate($\leq3<45\mu m$)  \\
2015 TB179 &LM-LAM-026   &Dhajala              &Rock         &Ordinary Chondrite &H3-4 Olivine-Bronzite           &Thin Section                \\
2015 TW144 &LM-LAM-026   &Dhajala              &Rock         &Ordinary Chondrite &H3-4 Olivine-Bronzite           &Thin Section                \\
\hline
\end{tabular}    
\end{table*}

\begin{figure*}[!htb]
\begin{center}
\includegraphics[width=5.7cm]{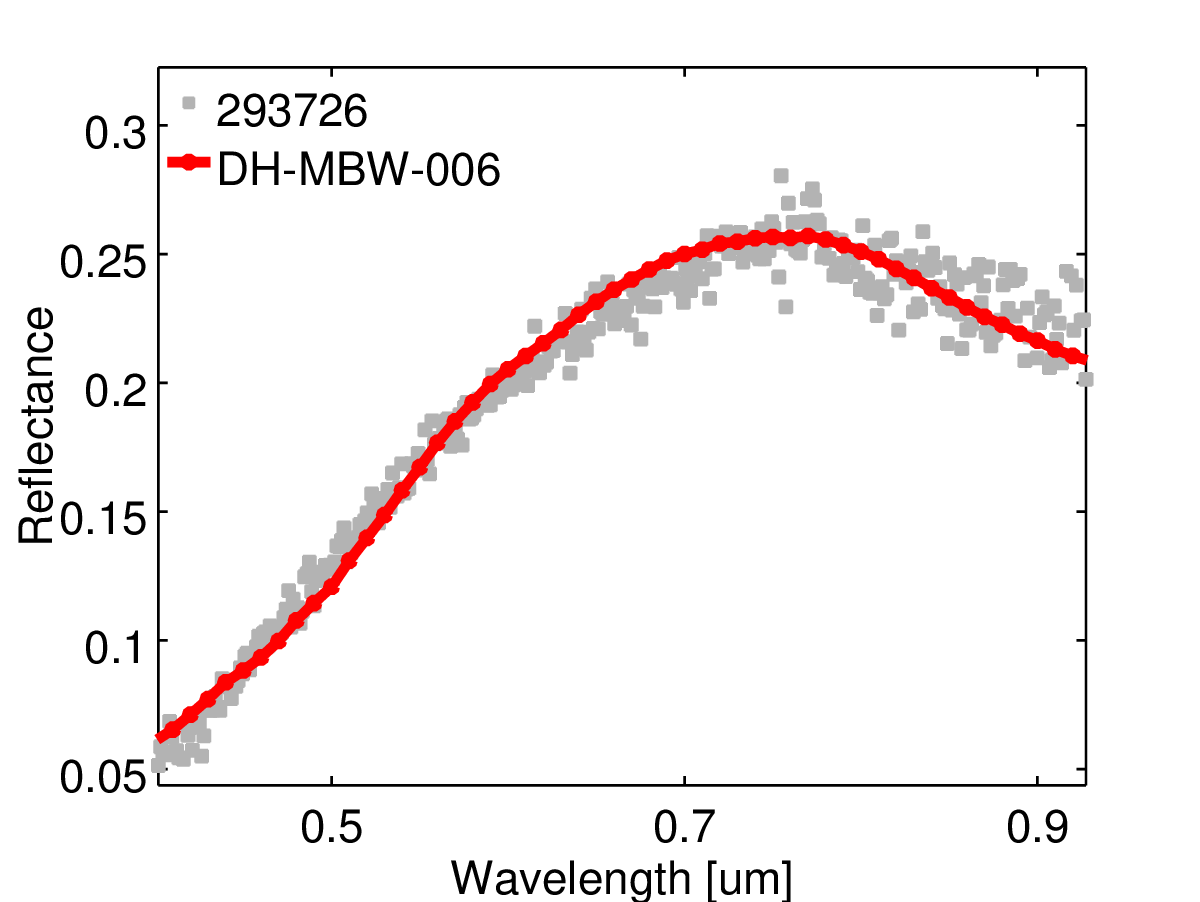}
\includegraphics[width=5.7cm]{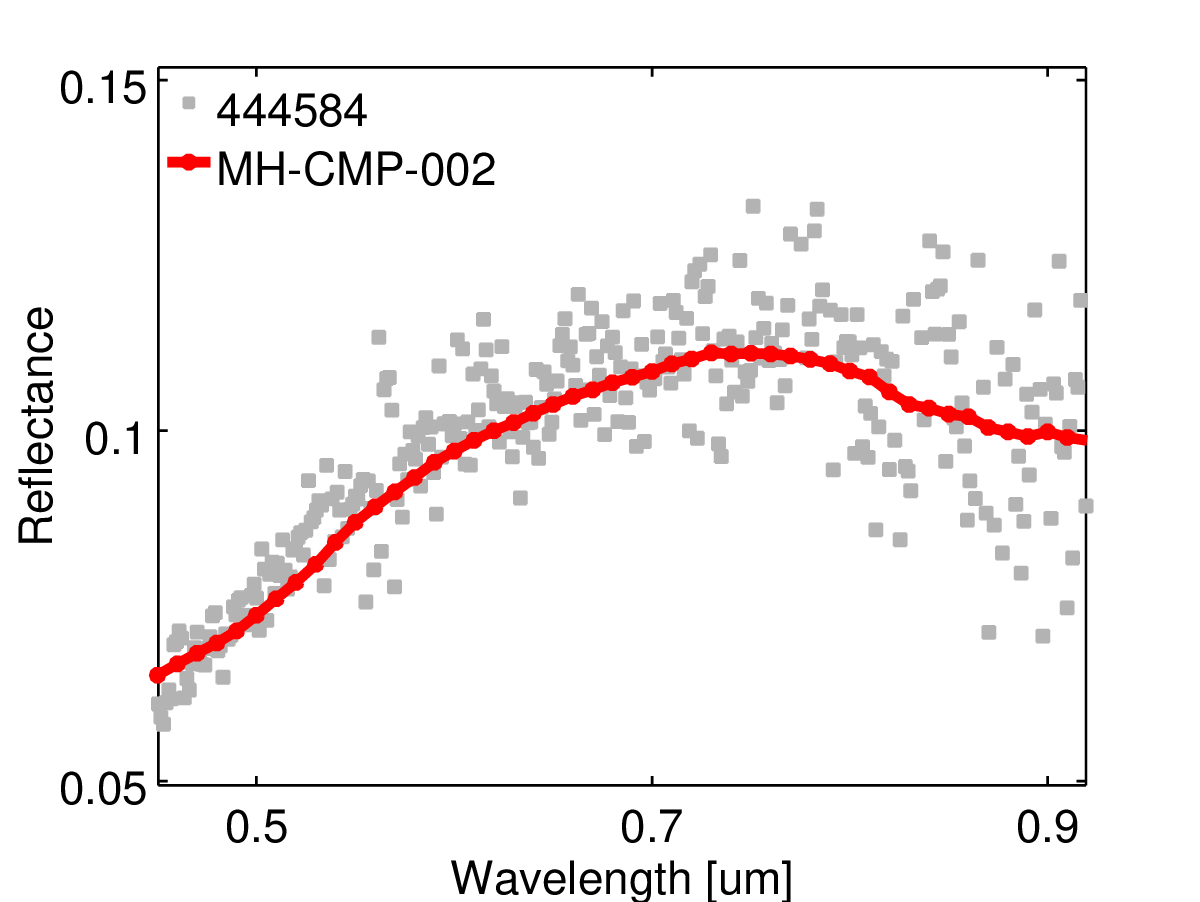}
\includegraphics[width=5.7cm]{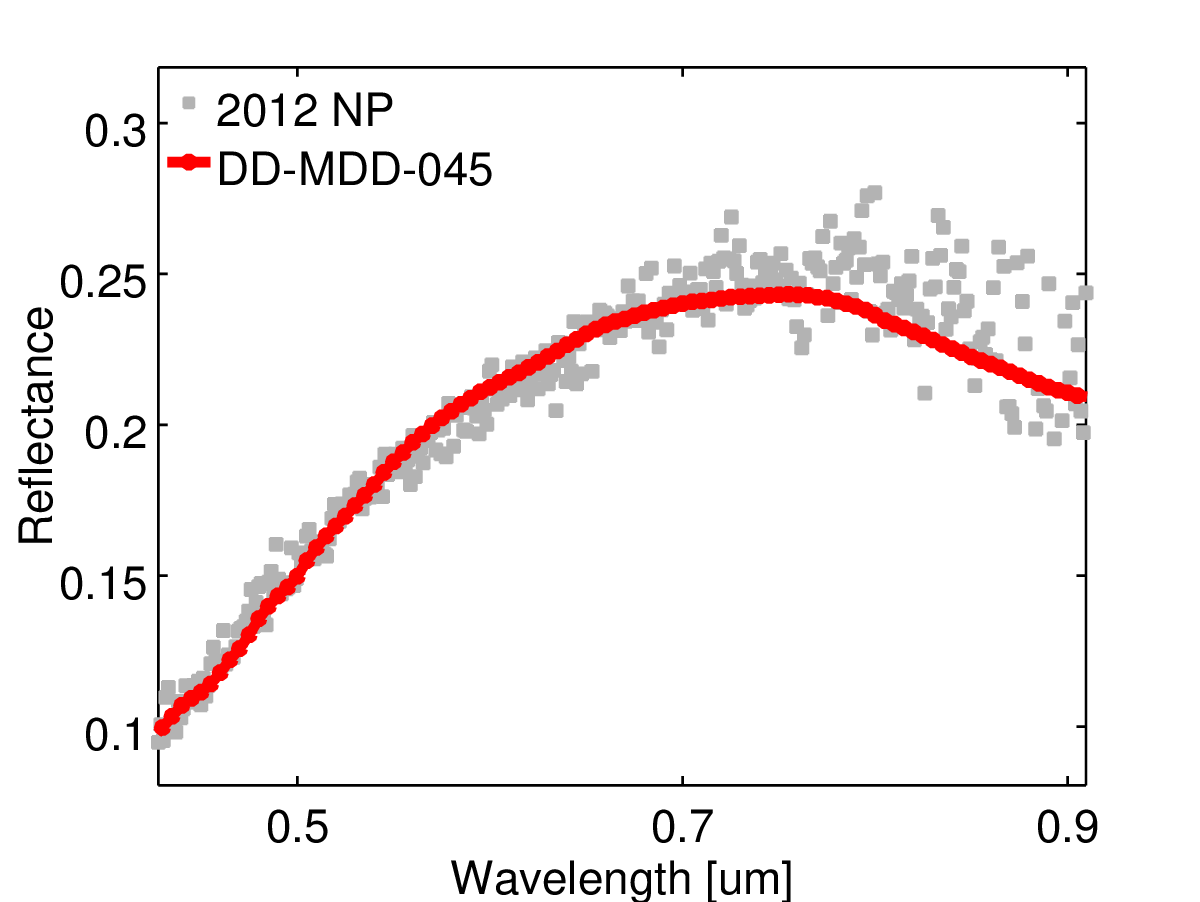}
\includegraphics[width=5.7cm]{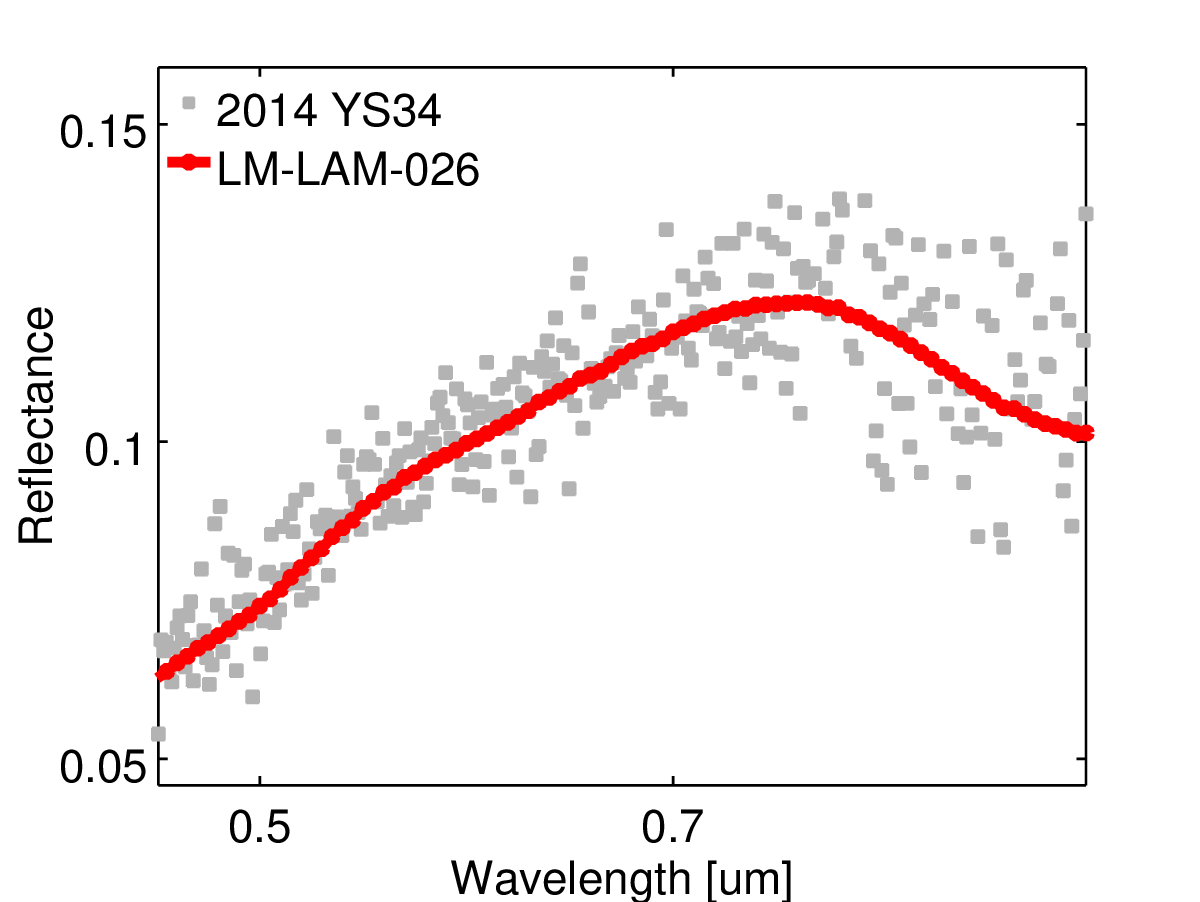}
\includegraphics[width=5.7cm]{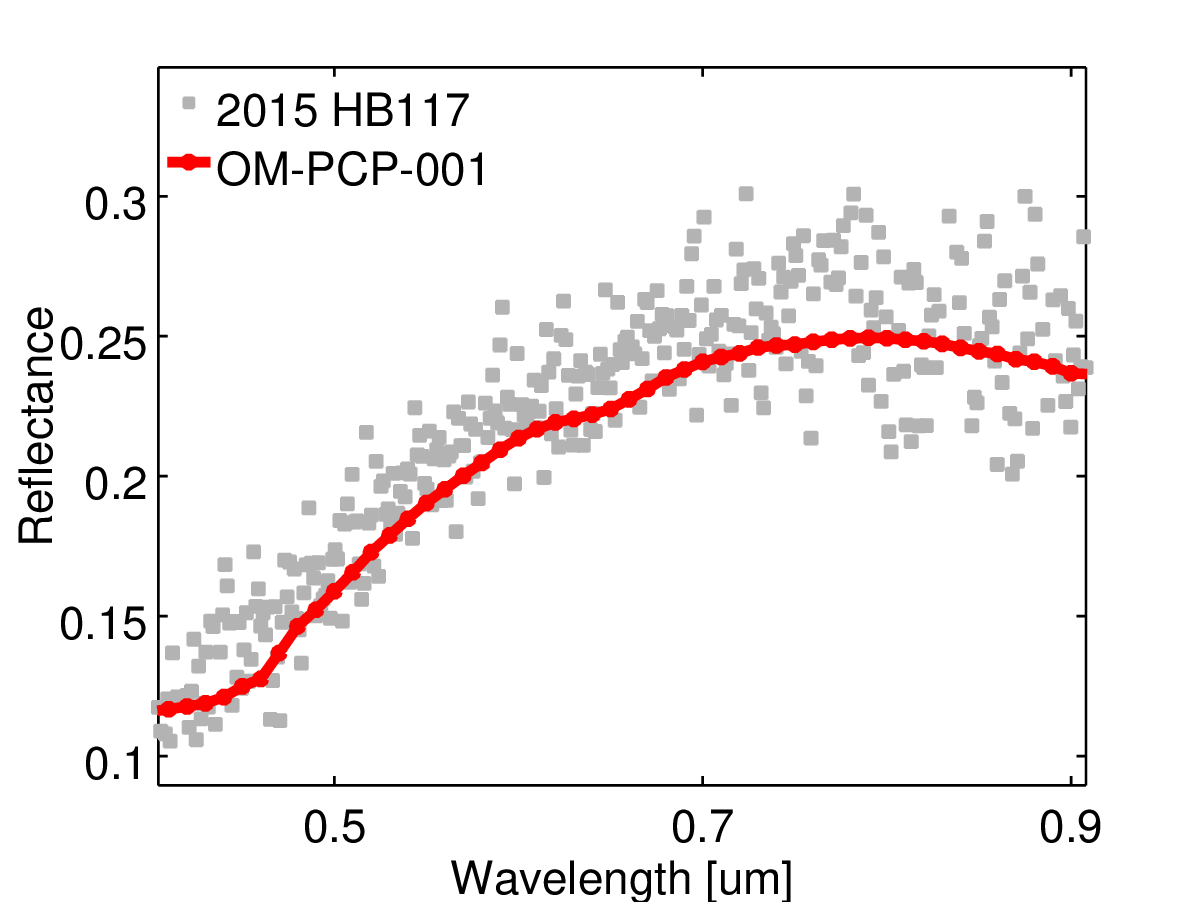}
\includegraphics[width=5.7cm]{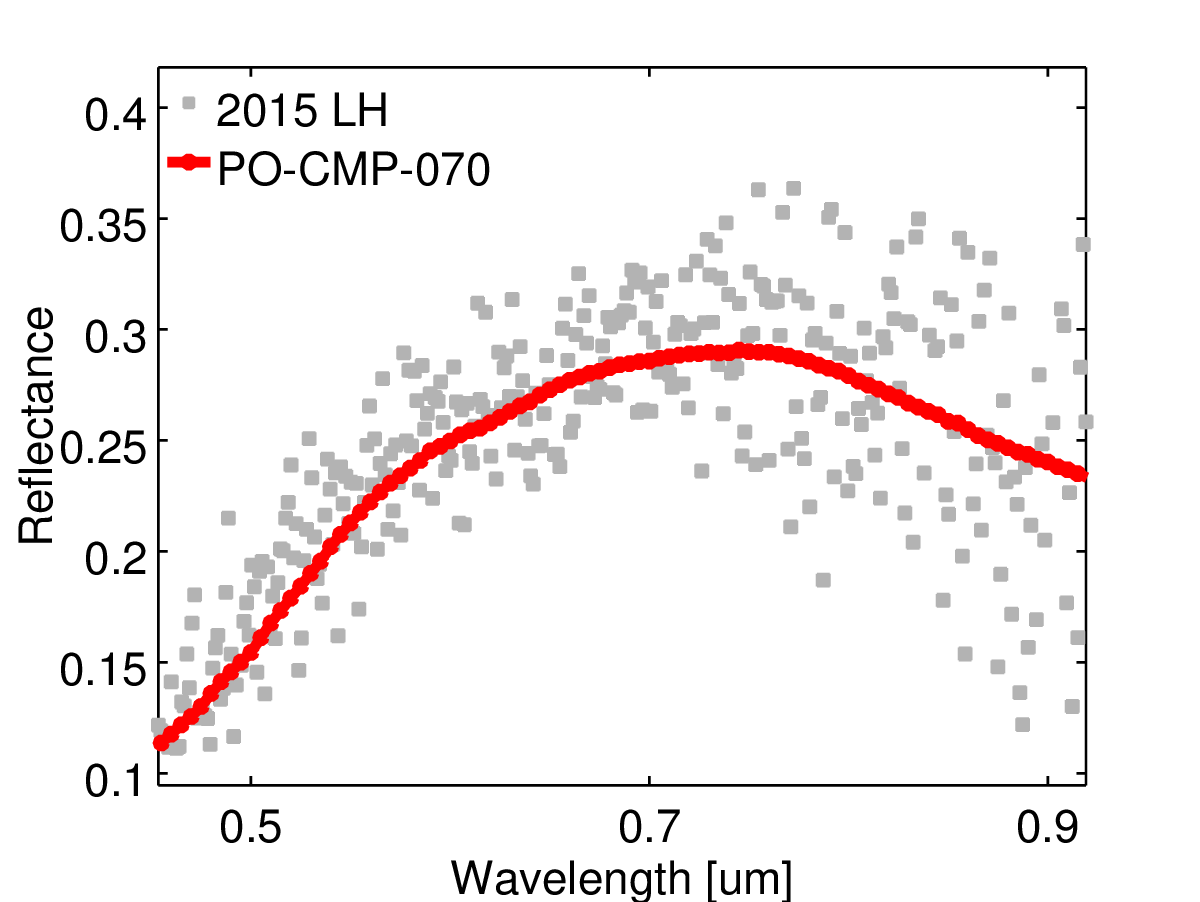}
\includegraphics[width=5.7cm]{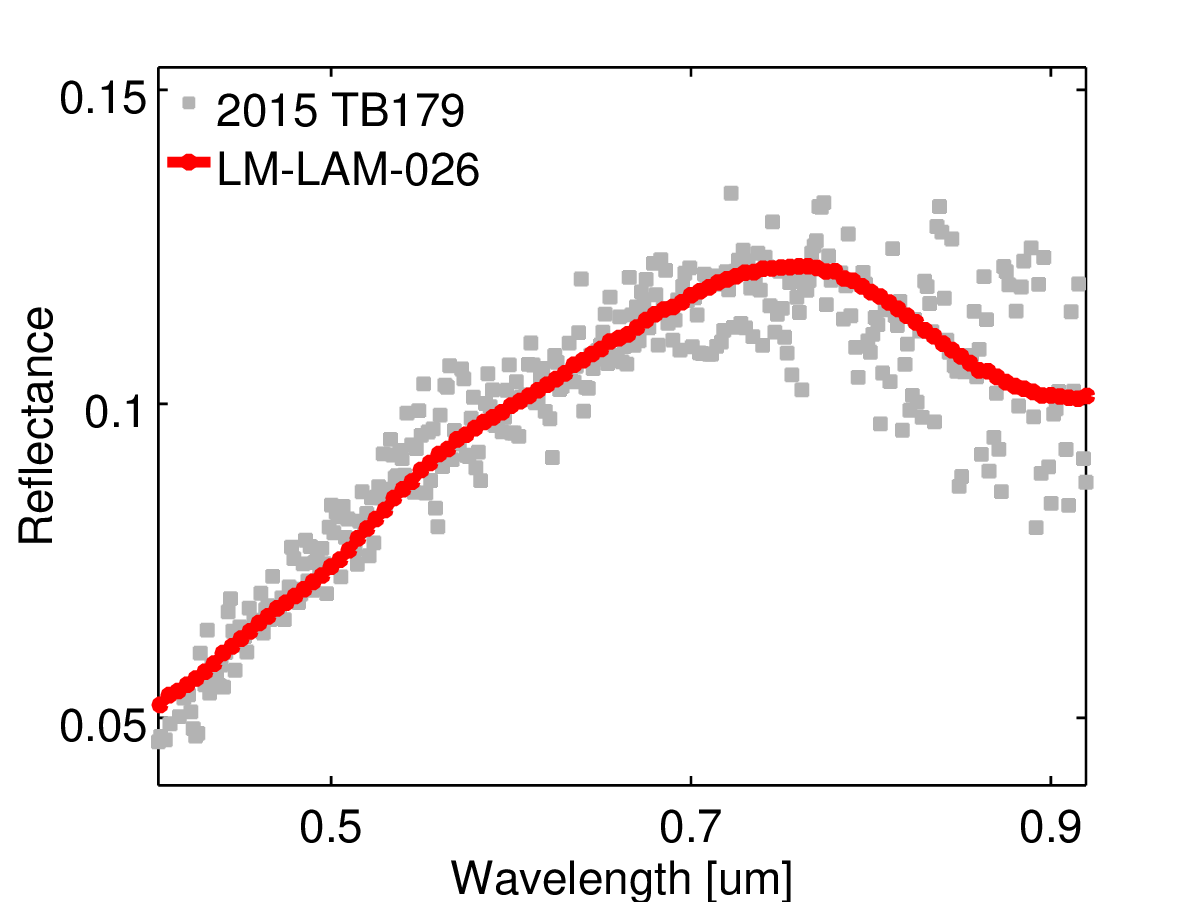}
\includegraphics[width=5.7cm]{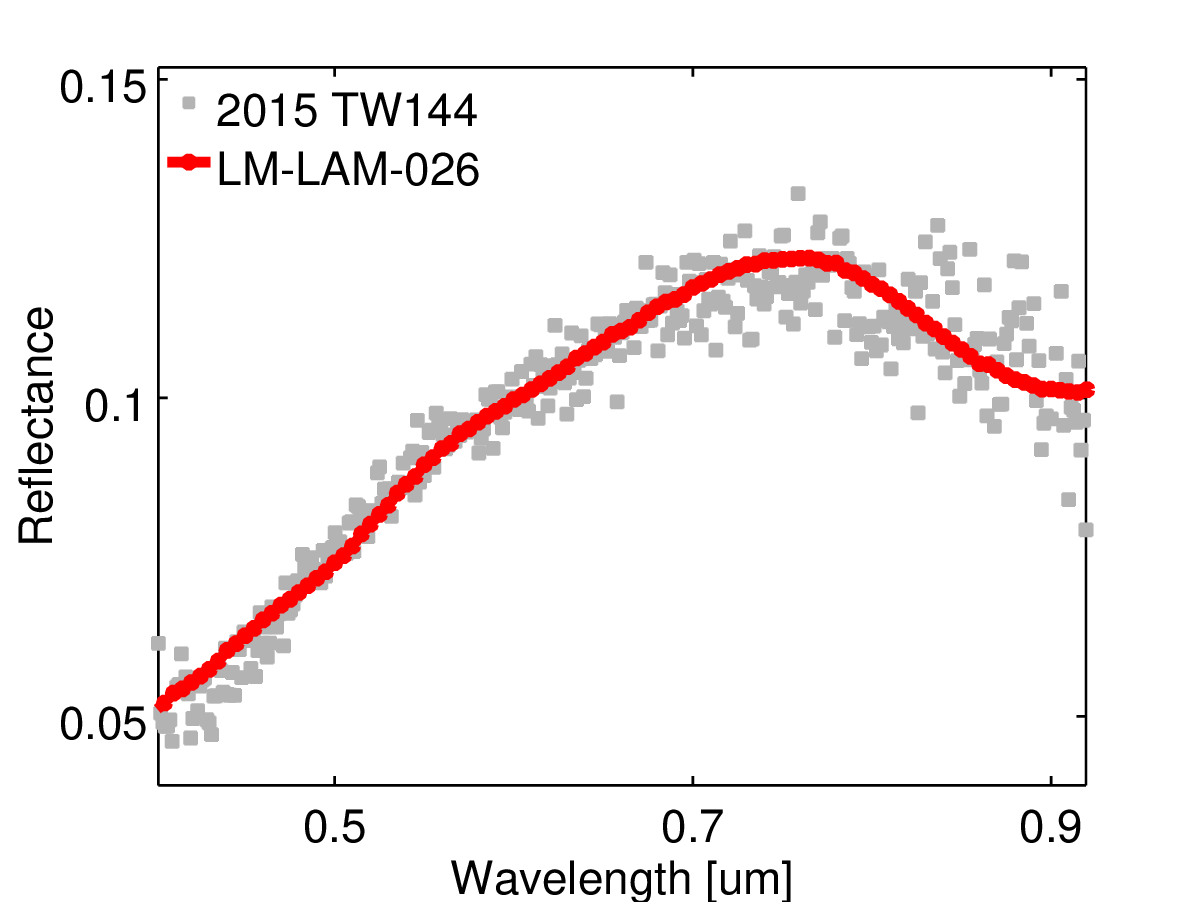}

\end{center}
\caption{Plot of the asteroid spectra versus the best match spectrum from the RELAB database. The asteroid is identified by its designation and it is plotted in grey, the RELAB spectrum is shown in red and is referenced by its sample ID (see Tabel~\ref{tab:RelabComparison} for additional details).}
\label{fig:Relab}
\end{figure*}

\newpage
\section{Spectral comparison with average spectra from Bus taxonomy}

\begin{figure*}[H]
\begin{center}
\includegraphics[width=5.7cm]{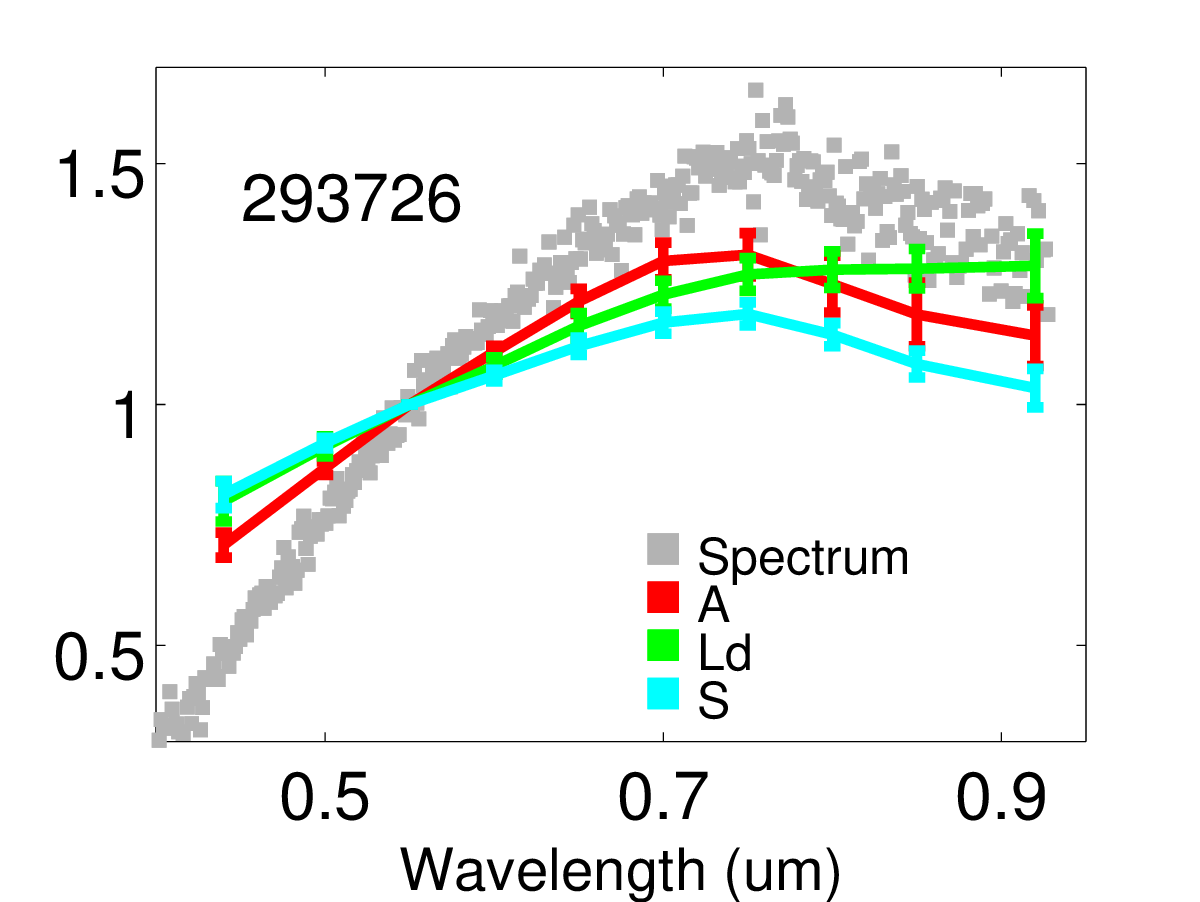}
\includegraphics[width=5.7cm]{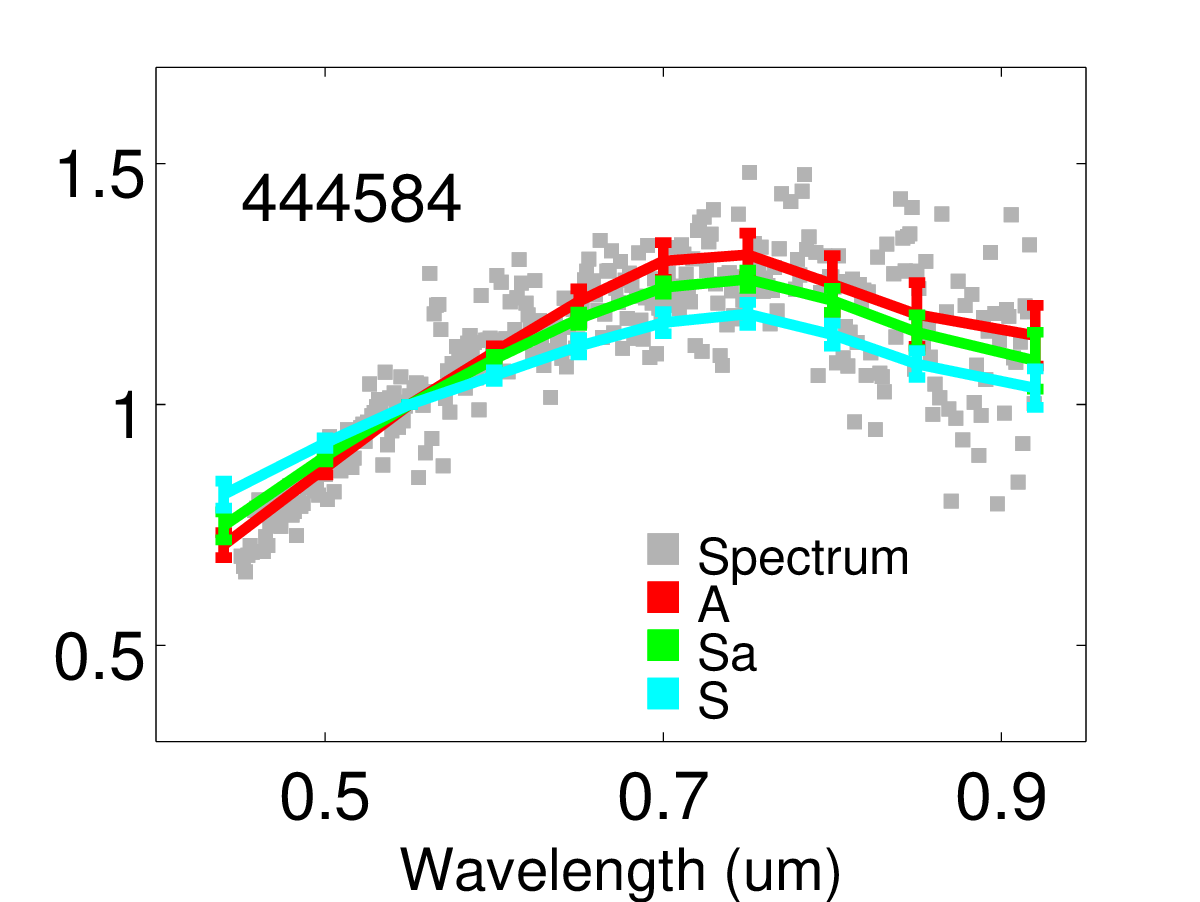}
\includegraphics[width=5.7cm]{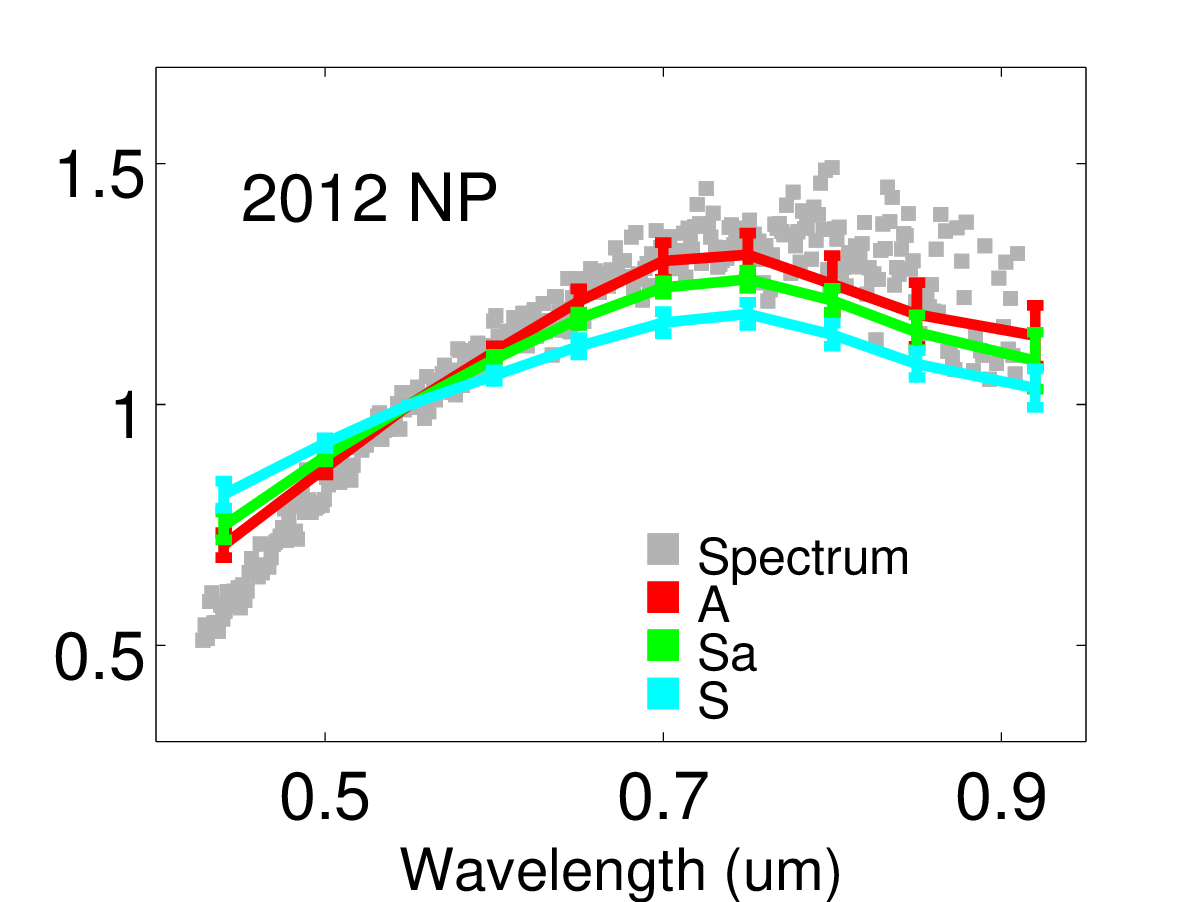}
\includegraphics[width=5.7cm]{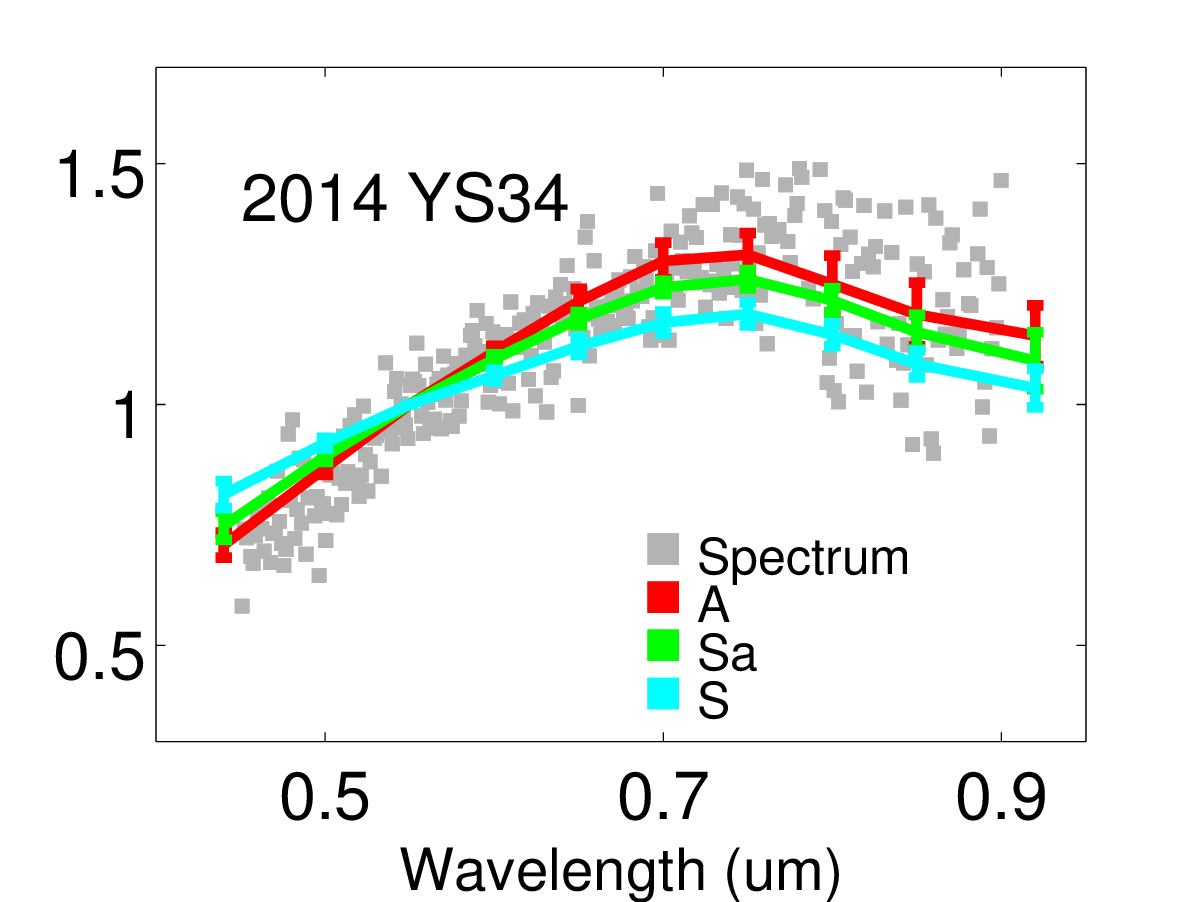}
\includegraphics[width=5.7cm]{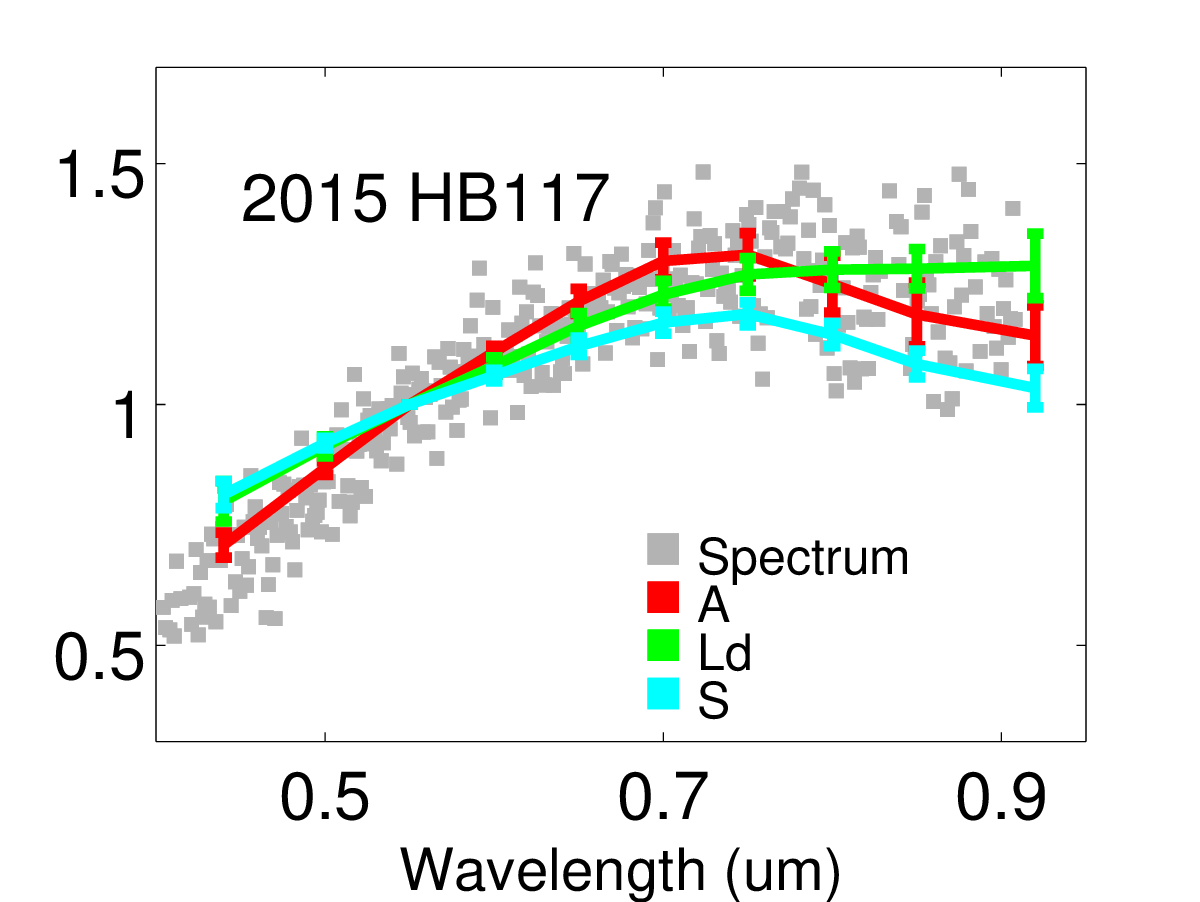}
\includegraphics[width=5.7cm]{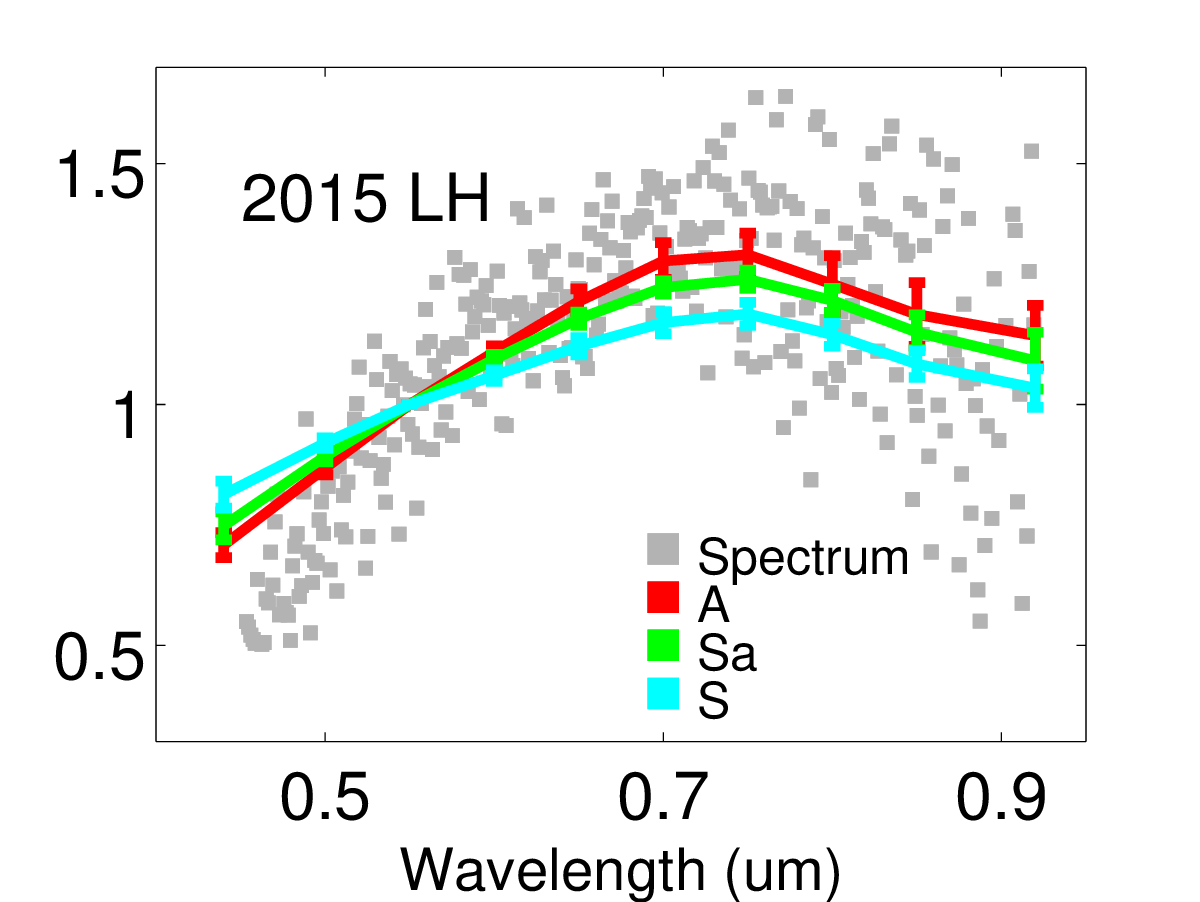}
\includegraphics[width=5.7cm]{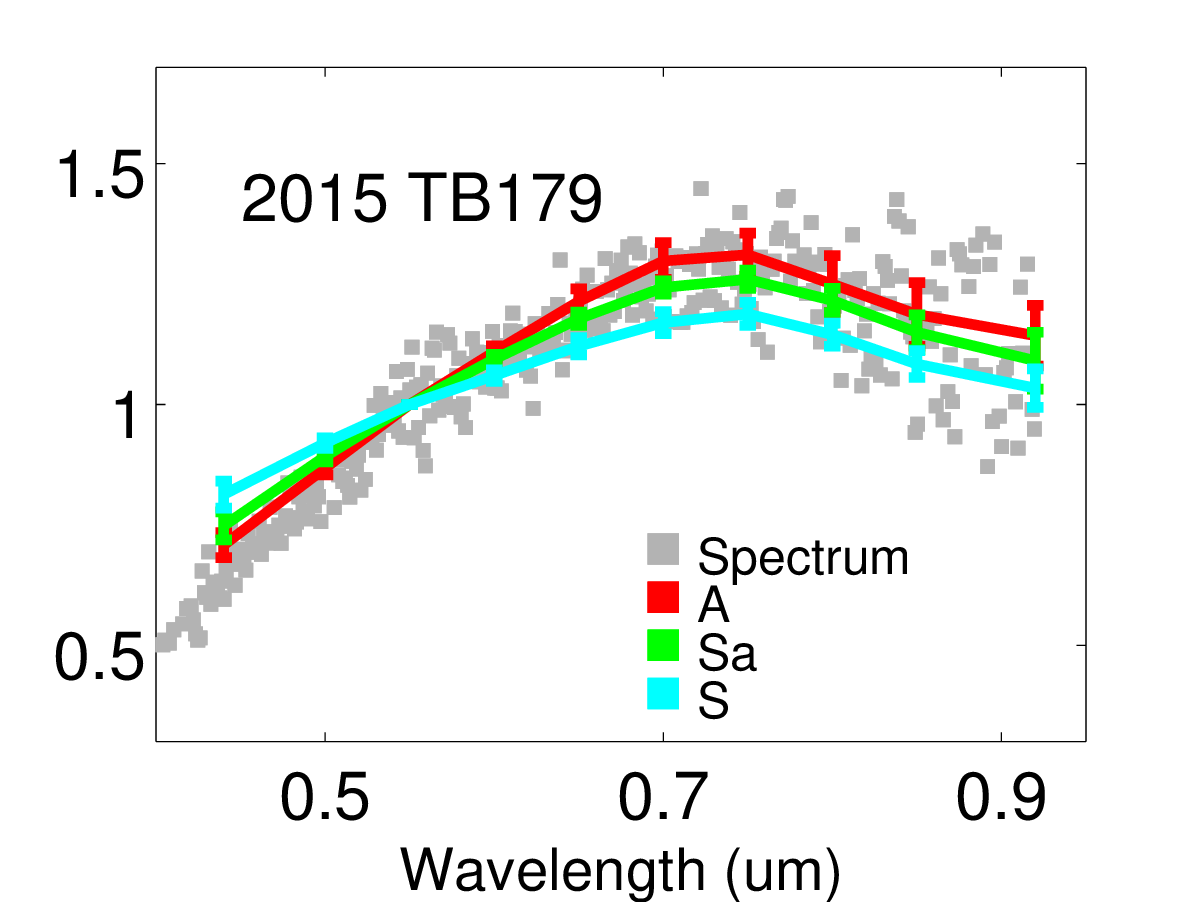}
\includegraphics[width=5.7cm]{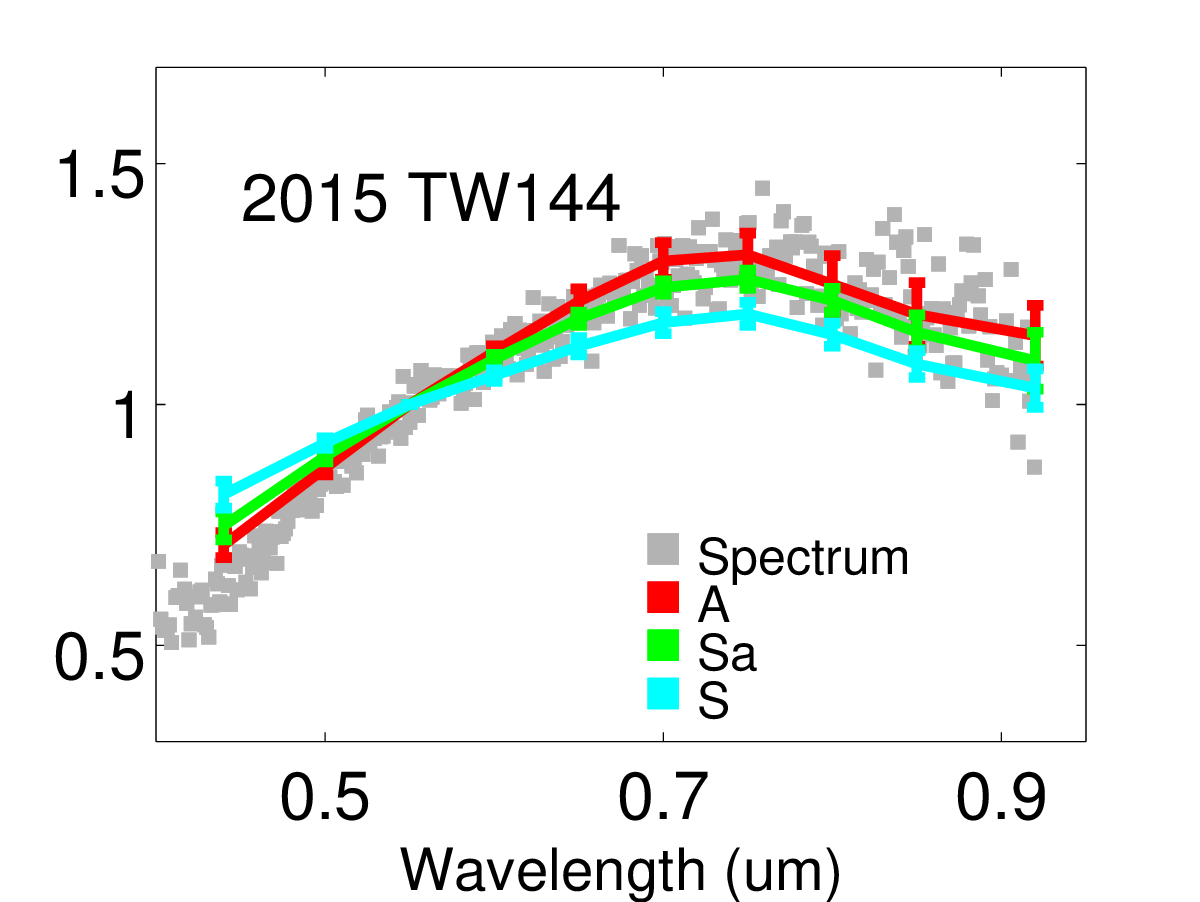}

\end{center}
\caption{Matching of the observed spectra with taxonomic templates defined by \citet{2002Icar..158..146B}. The visible spectra of (293726) 2007 RQ17, (444584) 2006 UK, 2012 NP, 2014 YS34, 2015 HB117, 2015 LH, 2015 TB179, 2015 TW144 are shown in grey. All spectra are normalized to 0.55 $\mu m$. The first two best matches reported in Table~\ref{tab:TaxMSq} are shown. The S-type which is the most common type in the near-Earth asteroid population is plotted for comparison.}
\label{fig:SpectraTaxa}
\end{figure*}


\bsp	
\label{lastpage}
\end{document}